\documentclass[11pt,oneside,letterpaper]{article}
\usepackage{amssymb}
\usepackage{amsmath}
\usepackage[dvips]{graphicx}
\usepackage{setspace}
\usepackage{amsfonts}
\usepackage{fancyhdr}
\usepackage{xcolor}
\usepackage{graphicx}
\usepackage{rotating}
\usepackage{comment}
\usepackage{color}
\usepackage{cite}
\usepackage{braket}

\usepackage{wrapfig}

\definecolor{darkgreen}{rgb}{0,0.5,0}
\definecolor{darkblue}{rgb}{0,0,0.6}
\definecolor{purple}{rgb}{0.4,.2,0.7}
\newcommand{\dd}{\partial}

\newcommand{\f}{\frac}

\newcommand{\be}{\begin{equation}}
\newcommand{\ee}{\end{equation}}

\usepackage[colorlinks=true,citecolor=darkgreen,linkcolor=black,urlcolor=purple]{hyperref}

\usepackage{pdfsync}

\makeatletter
\newcommand*{\defeq}{\mathrel{\rlap{%
                     \raisebox{0.3ex}{$\m@th\cdot$}}%
                     \raisebox{-0.3ex}{$\m@th\cdot$}}%
                     =}
\makeatother

\newcommand{\bea}{\begin{eqnarray}}
\newcommand{\eea}{\end{eqnarray}}
\def\be{\begin{equation}}
\def\ee{\end{equation}}

\let\a=\alpha \let\b=\beta \let\g=\gamma  
   
    \let\r=v
\let\s=\sigma \let\t=\tau  \let\c=\chi

    \let\L=\Lambda \let \S = \Sigma

\def\nn{\nonumber}
\let\f=\frac

\let\p=\partial
\def\be{\begin{eqnarray}}
\def\ee{\end{eqnarray}}
\def\ba{\begin{array}}
\def\ea{\end{array}}

\newcommand{\RR}{\mathbb{R}}

\newcommand{\mI}{\mathcal{I}}

\DeclareMathOperator{\Ext}{Ext}

\newcommand{\mS}{\mathcal{S}}

\allowdisplaybreaks  % allow page breaks in displayed eqs

\renewcommand*{\defeq}{\mathrel{\vcenter{\baselineskip0.5ex \lineskiplimit0pt
                     \hbox{\scriptsize.}\hbox{\scriptsize.}}}%
                     =}

\numberwithin{equation}{section}
\begin{document}
\onehalfspacing

\begin{center}

\vskip5mm

{\LARGE  {
Submerging islands through thermalization
 \\ \ \\ 
}}

Vijay Balasubramanian,$^{1,2}$ Ben Craps,$^{2}$ Mikhail Khramtsov,$^{3}$ and Edgar Shaghoulian$^{1}$

\vskip5mm
{\it $^1$ David Rittenhouse Laboratory, University of Pennsylvania,\\
  Philadelphia, PA 19104, USA}  \\
{\it $^2$ Theoretische Natuurkunde, Vrije Universiteit Brussel (VUB) and\\
The International Solvay Institutes, Brussels, Belgium }  \\
{\it $^3$ Department of Mathematical Methods for Quantum Technologies, \\ Steklov Mathematical Institute of Russian Academy of Sciences,\\ Gubkin str. 8, 119991 Moscow, Russia
}
\vskip5mm

{\tt vijay@physics.upenn.edu, Ben.Craps@vub.be, khramtsov@mi-ras.ru, eshag@sas.upenn.edu }

\end{center}

\vspace{4mm}

\begin{abstract}
\noindent
We illustrate scenarios in which Hawking radiation collected in finite regions of a reservoir provides temporary access to the interior of black holes through transient entanglement ``islands.''  Whether these islands appear and the amount of  time for which they dominate -- sometimes giving way to a thermalization transition -- is controlled by the amount of radiation we probe. In the first scenario,  two reservoirs are coupled to an eternal black hole. The second scenario involves two holographic quantum gravitating systems at different temperatures interacting through a Rindler-like reservoir, which acts as a heat engine maintaining thermal equilibrium. The latter situation, which has an intricate phase structure, describes two eternal black holes radiating into each other through a shared reservoir.

 \end{abstract}

\pagebreak
\pagestyle{plain}

\setcounter{tocdepth}{2}
{}
\vfill

\ \vspace{-2cm}
\renewcommand{\baselinestretch}{1}\small
\tableofcontents
\renewcommand{\baselinestretch}{1.15}\normalsize

%%%%%%%%%%%%%%%%%%%%%%%%%%%%%%%%%%%%%%%%%%%%%%%%%%%%%%%%%%%%%%%%%%%%%%%%
\section{Introduction}\label{intro}

The fine-grained entropy of any quantum system $A$  entangled with its complement $\bar{A}$ satisfies the unitarity bound $S(A) \leq$ min$\{\log \dim \mathcal{H}_A, \log \dim \mathcal{H}_{\bar{A}}\}$ in terms of the dimensions of the corresponding Hilbert spaces.  In a holographic theory, the entropy may be geometrized in terms of the areas of extremal surfaces in  spacetime \cite{RT,HRT,QES,Penington191,Almheiri191,Almheiri192}.   As the  system evolves,  entanglement structure can change dynamically, and be reflected in exchange of dominance between different extremal surfaces.  Beautiful work has shown that this exchange can be necessary for the satisfaction of the bound on $S(A)$ as time evolves \cite{Penington191,Almheiri191,Almheiri192,Almheiri193,Almheiri194,Almheiri195,Chen19,Penington192,MarolfMaxfield,HM,Rozali19,Bousso19,Balasubramanian:2020hfs,Hartman20}.

One example \cite{HM} involves  holographic CFTs entangled in a thermofield double state dual to eternal black holes connected  behind their horizons by a wormhole. The  entropy of a pair of subregions in these theories is initially associated to an extremal surface passing through the wormhole. The area of this  surface grows in time, threatening violation of the entropy bound, a fate  avoided by a thermalization transition after which the two regions no longer share mutual information.  After this time, their entropy is geometrized by a pair of disconnected extremal surfaces of constant area outside the horizons. A second example occurs in the same setup of eternal black holes, this time coupled to reservoirs collecting the Hawking radiation escaping the black holes \cite{Penington191,Almheiri191, Almheiri192, Almheiri194,Penington192}. In this case, at early times the entropy of the radiation increases in time exactly as computed by Hawking. At the Page time, a nontrivial quantum extremal surface (QES) \cite{RT, HRT, FLM, QES} appears in the spacetime. This  leads to a saturation of the entropy by requiring us to include the interior Hawking modes in the computation of the entropy. The nontrivial QES occurs due to the replica wormhole saddle points in the quantum gravity path integral \cite{Penington192,Almheiri194}. It bounds an ``island'' and is responsible for  information recovery from Hawking radiation via  access to the black hole interior. 
 
In this paper, we  study what occurs when one considers finite-sized portions of the radiation. For sufficiently small portions, the radiation may thermalize before or after it has a chance to encode the black hole interior. We will probe this competition between thermalization  and the island mechanism -- realizing both examples discussed above in the same physical system -- by examining the time-dependence of entanglement entropy. The various transitions can be predicted by a careful application of the unitarity bound described above. In short, island regions will appear when the entropy of quantum fields on the black hole background threatens the unitarity bound set by its Bekenstein-Hawking entropy, whereas thermalization will occur when the entropy of the radiation approaches the maximal amount dictated by the size of the radiation region's Hilbert space. We will exhibit a nontrivial phase structure transitioning between these possibilities in two scenarios, which we  now summarize.

\subsection*{\it{Summary of results}} 
Our first scenario, studied in section~\ref{sec:1xBH}, involves finite regions in two reservoirs coupled to a thermofield double black hole in AdS$_2$, see Fig.~\ref{fig:1BH} (the entropy of infinite radiation regions in this model was studied in \cite{Almheiri193,Almheiri194}).  The entropy initially grows linearly as time evolves, then (depending on the region size and boundary conditions) there may be a transition to half the initial rate, and finally the growth stabilizes; see Fig.~\ref{fig:PageCurve1BH}. If we treat the radiation as holographic in its own right, the transition between the initial and the final phase involves an exchange between an extremal surface passing between the reservoirs through the induced dimension, and a pair of disconnected extremal surfaces of constant area, as in \cite{HM}; see Fig.~\ref{fig:1BH-geod}. However the ``island formula'' (reviewed below) dictates a new intermediate phase, which includes an island, although the entropy does not stabilize but instead grows with only half the initial slope. The eventual stabilization of entanglement entropy of finite intervals then happens not because of entanglement islands, but because of the thermalization of the segments. When the entanglement entropy saturates, the island is no longer accessible. 

\begin{figure}[t]
	\centering
	\includegraphics[scale=0.5]{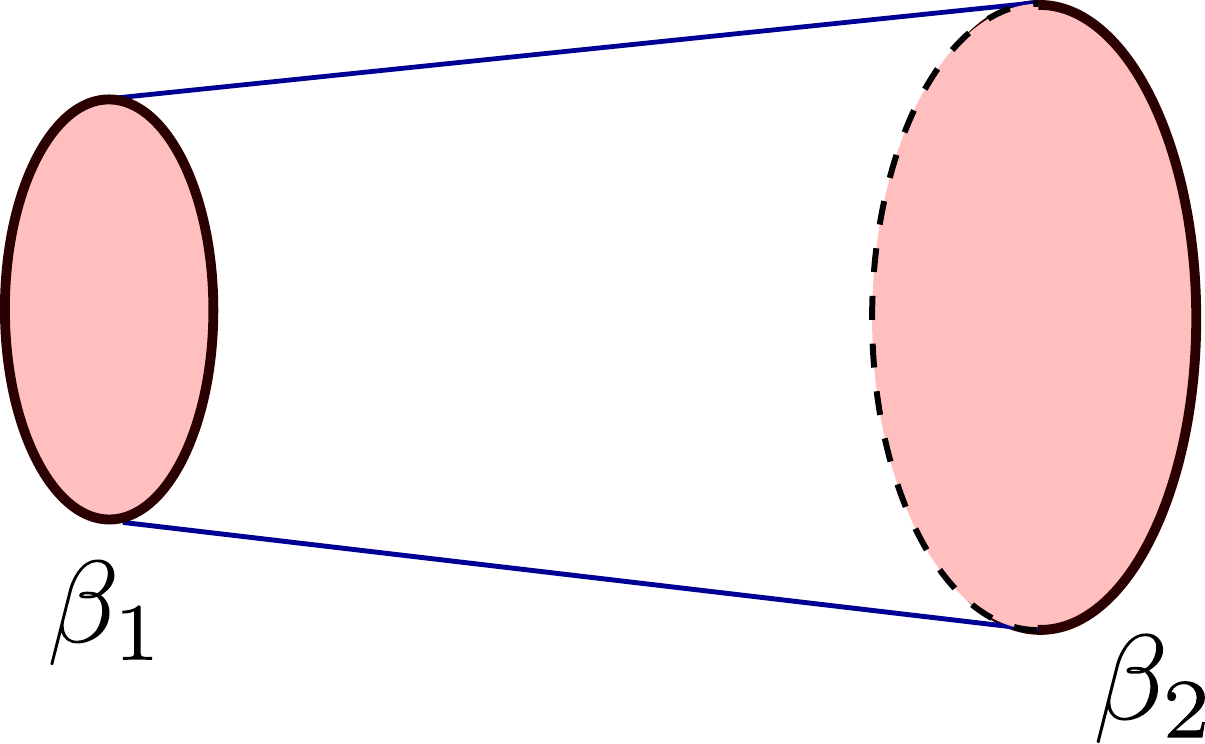}
	\caption{The schematic picture of the black hole phase in the Euclidean signature. The spacetime is two hyperbolic disks connected by the conical reservoir. }
	\label{fig:w2isks}
\end{figure}

Our second scenario involves two holographic quantum dots at different temperatures, each dual to an eternal AdS$_2$  black hole, in local equilibrium with a finite radiation reservoir. The reservoir is modeled by Rindler space, which acts as a heat engine  maintaining local equilibrium by redshifting warmer modes approaching the cooler black hole and vice versa; see the beginning of section~\ref{sec3} and Fig.~\ref{fig:SYK-bCFT-SYK}.  The radiation theory will itself be holographic.  The phase structure includes transitions between an asymmetric wormhole (the ``confined" phase; see section \ref{sec:WHphase} and Fig.~\ref{fig:WH}) and two black holes of different temperature (the ``deconfined" phase; see section \ref{sec:BHphase} and Fig. \ref{fig:w2isks}). In the deconfined phase, many extremal surfaces vie to dominate the entropy of reservoir regions in this model. This is studied in section \ref{sec4}; see Fig.~\ref{fig:EE-2sided+BH}, \ref{fig:EE-2sided+BH-geod}, \ref{fig:Page-1+1BH} for the case where we include one pair of thermofield double quantum dots in the region whose entropy we are computing and  Fig.~\ref{fig:EE-2sided-bath}, \ref{fig:EE-2sided-bath-geod}, \ref{fig:Page-bath-equal} for the case where we do not. As time evolves, these surfaces exchange dominance to maintain the unitary upper bound on entropy. In particular, unlike the single black hole case, in this model it is possible to have both temporary island configurations and permanent island configurations.

\section{An eternal black hole coupled to a reservoir}
\label{sec:1xBH}

\begin{figure}[t]
\begin{center}
\includegraphics[scale=0.8]{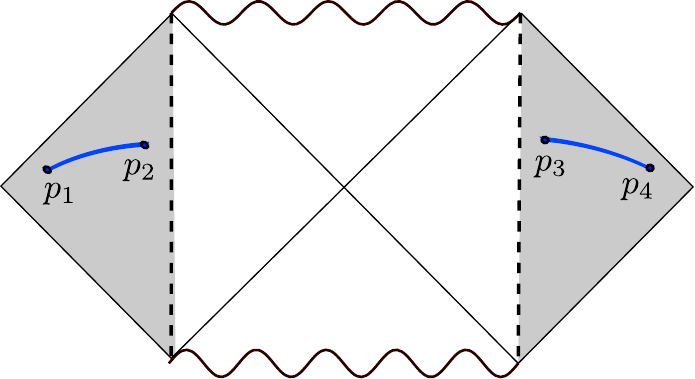} 
\caption{Penrose diagram of an eternal AdS$_2$  black hole with flat reservoirs glued to the boundaries, and identical equal-time segments $[p_1, p_2]$ and $[p_3, p_4]$. We compute the entanglement entropy of the union of these segments.
}
\label{fig:1BH}
\end{center}
\end{figure}

Consider an AdS$_2$ eternal black hole with flat, non-gravitating radiation reservoirs glued to the two boundaries (Fig.~\ref{fig:1BH}) \cite{Almheiri193,Almheiri194}.  In the AdS$_2$ region we have Jackiw-Teiltelboim (JT) gravity, along with transparent boundary conditions for conformal matter in the black hole and reservoirs.  The action is:
\be\label{JTact}
I = - \f{\phi_0}{4\pi}\left[\int_{\S_2} R + 2 \int_{\p\S_2} K \right] -\f{1}{4\pi} \left[\int_{\S_2} \phi (R+2) + 2\phi_b \int_{\p \S_2} (K-1) \right] + I_{\text{CFT}}\,,
\ee
where the first term is topological, the last term is a Conformal Field Theory (CFT) with central charge $c$, $\phi_0$ gives the Bekenstein-Hawking entropy of the extremal black hole (we have set $4 G_N = 1$), and $\phi_b$ is the asymptotic value of the dilaton on $\Sigma_2$, the region of spacetime where gravity is dynamical. Varying (\ref{JTact}) with respect to the dilaton fixes the metric to be locally AdS$_2$. Globally we  consider the eternal black hole (Fig.~\ref{fig:1BH}), with each of the two exterior regions described by 
\be
ds^2_{\text{grav}} = \f{4\pi^2}{\b^2}\f{-dt^2 + d\sigma^2}{\sinh^2 \f{2\pi \sigma}{\b}}\,, \qquad t \in \RR\,,\quad \sigma \in (-\infty, -\epsilon]\,. \label{BHmetsol}
\ee
We glue the surface $\sigma = -\epsilon$  continuously to the reservoir \cite{Almheiri193}, so the latter's metric is:
\be
ds^2_{\text{bath}} = \f{-dt^2 + d\sigma^2}{\epsilon^2}\,,\qquad t \in \RR\,,\quad \sigma \in [-\epsilon, +\infty)\,.
\ee
Varying with respect to the metric yields the dilaton equation of motion: 
\be
D_\mu D_\nu \phi - g_{\mu\nu} \Box \phi + g_{\mu\nu} \phi + 2\pi T_{\mu\nu} =0 \label{dilatonEOM}\,.
\ee
With stress tensor given by $T_{\mu\nu} = \frac{c}{24\pi} g_{\mu\nu}$, the dilaton in the gravitational region (\ref{BHmetsol}) is 
\be
\phi(\sigma) = \f{2\pi \phi_r}{\b} \coth \f{2\pi \sigma}{\b}\label{BHdilsol}\,,
\ee
where $\phi_r$ is an integration constant. 

\subsection{Radiation entropy}
\label{sec:1xBH-setup}

The island formula \cite{Penington191, Almheiri191, Almheiri192,Almheiri194,Penington192} says that the entanglement entropy of reservoir region $A$ is
\be
\mS(A) = \min\, \underset{\mI}{\text{ext}} \left[ S_{\text{CFT}}(A \cup \mI) + \text{Area}(\dd \mI)\right]\,, \label{IslandFormula}
\ee
where $\mI$ is an ``island''  in the gravitating region. In JT gravity the area term equals the dilaton value on the corresponding surface plus the constant $\phi_0$, while, for a general 2d CFT, the entropy on  $A \cup \mI$ is related to the (generically non-universal) correlator of twist operators. We take the CFT to be holographic following  \cite{Almheiri192}\footnote{For further developments of the doubly holographic approach to the entanglement entropy of radiation, see e.g. \cite{Rozali19,Sully20,Myers1,Myers2,Myers3,Chen:2020jvn,Geng:2020qvw,Geng20,Geng21}, especially \cite{Myers1} for a pedagogical treatment.}, and compute $ S_{\text{CFT}}(A \cup \mI)$ through the Ryu-Takayanagi formula \cite{RT,HRT} in a 3d gravity theory  dual to this CFT on a fixed curved background.

In Euclidean signature, the CFT on the 2d boundary of the 3d gravity theory is defined on a Euclidean black hole attached to reservoirs with fixed metrics. The dynamical part of the 2d boundary is referred to as a ``Planck brane'', which corresponds to a cut-off boundary of the 3d geometry. Given a specific dynamical 2d metric $ds_2^2$ and stress tensor, it is convenient to introduce  a complex coordinate $w$ on the 2d boundary so that $T_{ww}= 0$ in the flat Weyl-transformed metric $dwd\bar w$ \cite{Almheiri192}. Since the stress tensor vanishes,  the  dual 3d spacetime can be described by Poincar\'e coordinates in Euclidean AdS$_3$:
\be
ds^2 = \frac{dw d\bar{w} + dz^2}{z^2}\,. \label{Metric-AdS3-P-E}
\ee
Writing the original boundary metric (before Weyl transformation) as
\begin{equation}
ds_2^2 = \Omega^{-2}(w, \bar{w}) dw d\bar{w} \,, \label{Metric-Weyl-flat}
\end{equation}
the holographic relation
\begin{equation}
g_{\mu\nu}^{(3)}\large|_{\text{bdy}} = \frac{1}{\varepsilon^2}g_{\mu\nu}^{(2)} %\label{metric-condition}
\end{equation}
leads to
\be
\frac{dw d\bar{w}}{z(w)^2} = \frac{1}{\varepsilon^2} \Omega^{-2}(w, \bar{w}) dw d\bar{w} \, \,  \Rightarrow \, \, z(w) = \varepsilon \Omega(w, \bar{w})\,, \label{z(w)}
\ee
which determines the embedding of the Planck brane in the 3d geometry \cite{Almheiri192}. Note that the regulator $\varepsilon$ is distinct from the 2d cutoff $\epsilon$ where the reservoir is glued to the AdS$_2$ black hole. 

Following \cite{RT,HRT} the CFT entropies
$ S_{\text{CFT}}(A \cup \mI)$ are computed by  lengths of geodesics in (\ref{Metric-AdS3-P-E}) ending on boundary points of $A \cup \mI$,  with $z(w)$ treated as a cutoff. Physical quantities will depend nontrivially on $\Omega$, e.g.,  the entropy of a single  interval between $w_1$ and $w_2$ is \cite{CC,Fiola94}:
\be
S(w_1, w_2) = \frac{c}{6} \log \left(\frac{|w_1- w_2|^2}{\varepsilon_1 \varepsilon_2 \Omega(w_1, \bar{w}_1) \Omega(w_2, \bar{w}_2)}\right)\,, \label{EE-CFT}
\ee
To return to  Lorentzian coordinates (\ref{BHmetsol}), we write  $w = e^{\frac{2\pi}{\beta} (\sigma + i \tau)}$ and Wick rotate $\tau = i t$. The single interval entropy is universal for  2d CFTs \cite{CC} but we consider a union of intervals.  Typically, several combinations of geodesics end on such unions, with each geodesic length given by (\ref{EE-CFT}).  For fixed $A \cup \mI$ the minimal total geodesic length gives the entropy.  Thus, we identify possible 3d geodesic configurations computing  $ S_{\text{CFT}}(A \cup \mI)$, extremize the functional in (\ref{IslandFormula}) for each choice, and select the minimizing choice.

\subsection{The entanglement entropy of finite segments}

\begin{figure}[t]
\begin{center}
\includegraphics[scale=0.55]{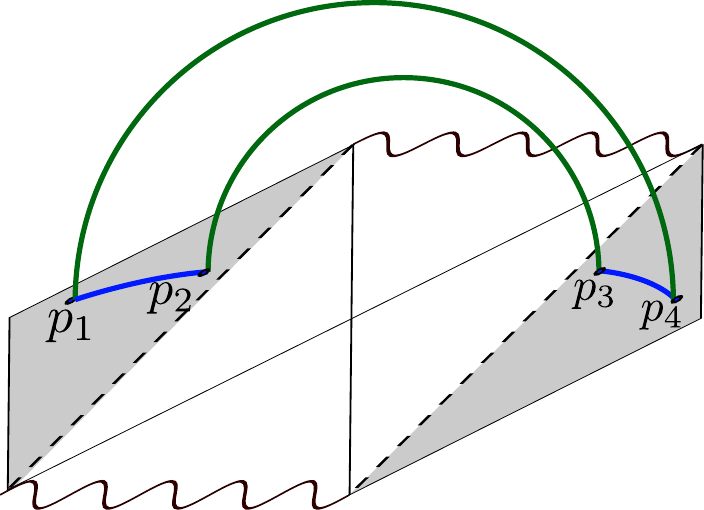} \hspace{2cm} 
\includegraphics[scale=0.6]{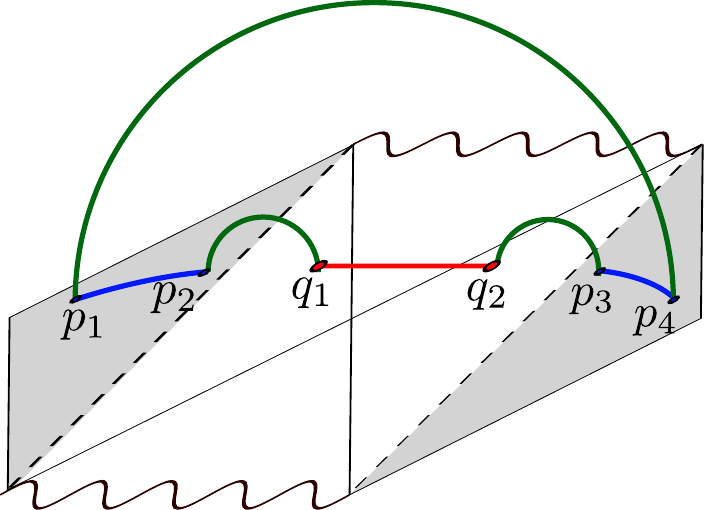} \\ 
(a) \hspace{5cm} (b) \\ 
\includegraphics[scale=0.6]{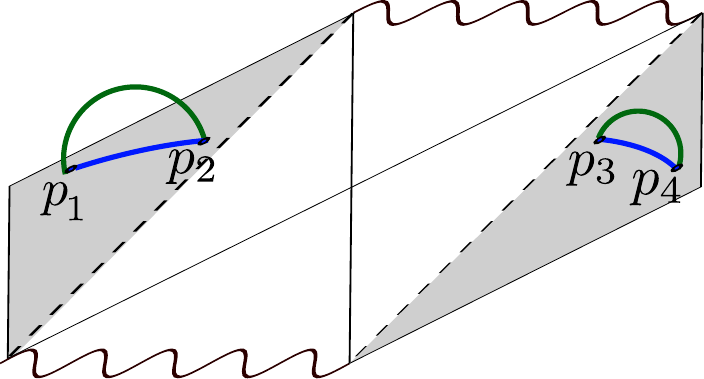} \hspace{2cm} 
\includegraphics[scale=0.6]{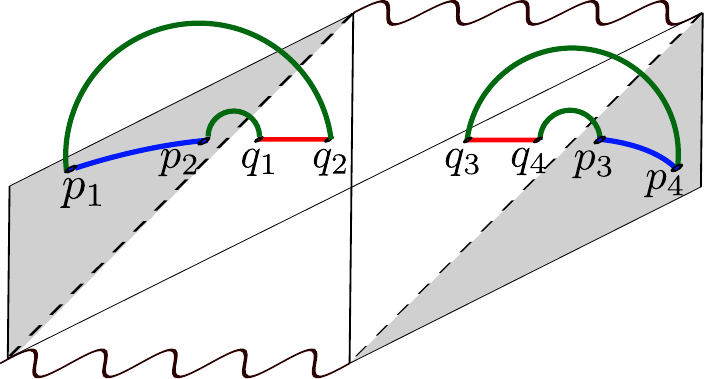} \\ 
(c) \hspace{5cm} (d) 
\caption{Extremal surfaces that extend into the 3d bulk from the Penrose diagram in Fig.~\ref{fig:1BH} playing the role of the 2D boundary. (a) $\rightarrow$ (c) is a  thermalization transition, which can be interrupted by the island configuration (b). Surface (d) is always subleading. }
\label{fig:1BH-geod}
\end{center}
\end{figure}

To compute the entropy of the region  $A = [p_1, p_2] \cup [p_3, p_4]$ in Fig.~\ref{fig:1BH}, we will use a time that continuously glues global AdS$_2$ to the reservoirs, reversing  Schwarzschild time $t$ (\ref{BHmetsol}) in one of the  exterior regions of the black hole. Thus we choose endpoint coordinates
\be
p_1 = \left(b, -t+i\frac{\beta}{2}\right); \quad p_2 = \left(a, -t+i\frac{\beta}{2}\right); \quad p_3 = (a, t); \quad p_4 = (b, t)\,.
\ee
Fig.~\ref{fig:1BH-geod} shows 3d geodesics between  interval endpoints $p_{1,2,3,4}$ and a possible island in the black hole region.\footnote{UV divergences are associated with endpoints of $p_{1,2,3,4}$ and possible islands. The first kind are the same for any choice of geodesic, so we omit them. The second kind renormalize  $\phi_0$.  }
The green curves are  Ryu-Takayanagi surfaces, while red regions inside the black hole  are  islands whose  endpoints are the quantum extremal surfaces found by extremizing (\ref{IslandFormula}).

\paragraph{Configuration (a): linear growth.} Two 3d geodesics connect  $p_1 \leftrightarrow p_4$ and $p_2 \leftrightarrow p_3$, respectively. %The  entropy  does not include  quantum extremal surfaces and is
There are no islands and the entropy is given by
\bea
\mS_a = S_{\text{connected}}^{\text{no island}} (p_1, p_2; p_3, p_4) = 2\,\f{c}{3}\log\left[\f{\pi}{\beta}  \cosh\f{2 \pi  t}{\b }\right]\,. \label{Sno-island}
\eea
This expression grows approximately linearly in time.

\paragraph{Configuration (b): island.} This fully connected configuration includes an island in the black hole region between $[q_{1}, q_{2}]$. The location is obtained by extremizing (\ref{IslandFormula}) with respect to $q_{1,2} = (x, t_x)$ in their respective black hole exterior patches. The entropy is
\bea
S_{\text{connected}}^{\text{island}} &=& 2\left(\phi_0 + \frac{2\pi \phi_r}{\beta} \coth \left(-\frac{2\pi}{\beta} x\right) \right) +\frac{c}{3} \log \left(\frac{\beta \left(\cosh \left[\frac{\pi}{\beta} (x- a)\right] - \cosh \left[ \frac{2\pi}{\beta} (t - t_x)\right]\right)}{\pi \sinh\left(-\frac{2\pi x}{\beta}\right)}\right) \nn\\
&+& \f{c}{3}\log\left[\f{\pi}{\beta}  \cosh\f{2 \pi  t}{\b }\right] \,.\label{Sisland}
\eea
The extrema for the time and space decouple, and the time equation yields $t_x = t$. The  solution for $x$ is cumbersome, but in the regime 
$\f{\phi_r}{c\b} \gg 1$ there is a simplified expression
\be
x \approx a + \f{\b}{2\pi} \log \left[24\pi \f{\phi_r}{c\b}\right]\,,\qquad \f{\phi_r}{c\b} \gg 1\,.
\ee
The first line in (\ref{Sisland}) determines the  time-independent island contribution. The second line, which grows linearly with half the slope compared to the configuration (a), comes from the long geodesic that goes across the Einstein-Rosen bridge. 

\paragraph{Configuration (c): thermalization.} This  configuration gives the sum of thermalized CFT entanglement entropies for the  thermofield double copies of the reservoir segment: 
\be
S_{\text{thermalized}} = 2\frac{c}{3} \log\left( \frac{\beta }{\pi} \sinh \frac{\pi |a - b|}{\beta}\right) \,. \label{Stherm}
\ee

\paragraph{Configurations (d): disconnected islands.} Here two islands lie in exterior regions of the black hole, connected with the RT geodesics to their respective copies of the radiation segment. One can show that this configuration is always subleading compared to (c) while also being time independent.

\begin{figure}[t]
\begin{center}
\includegraphics[scale=0.45]{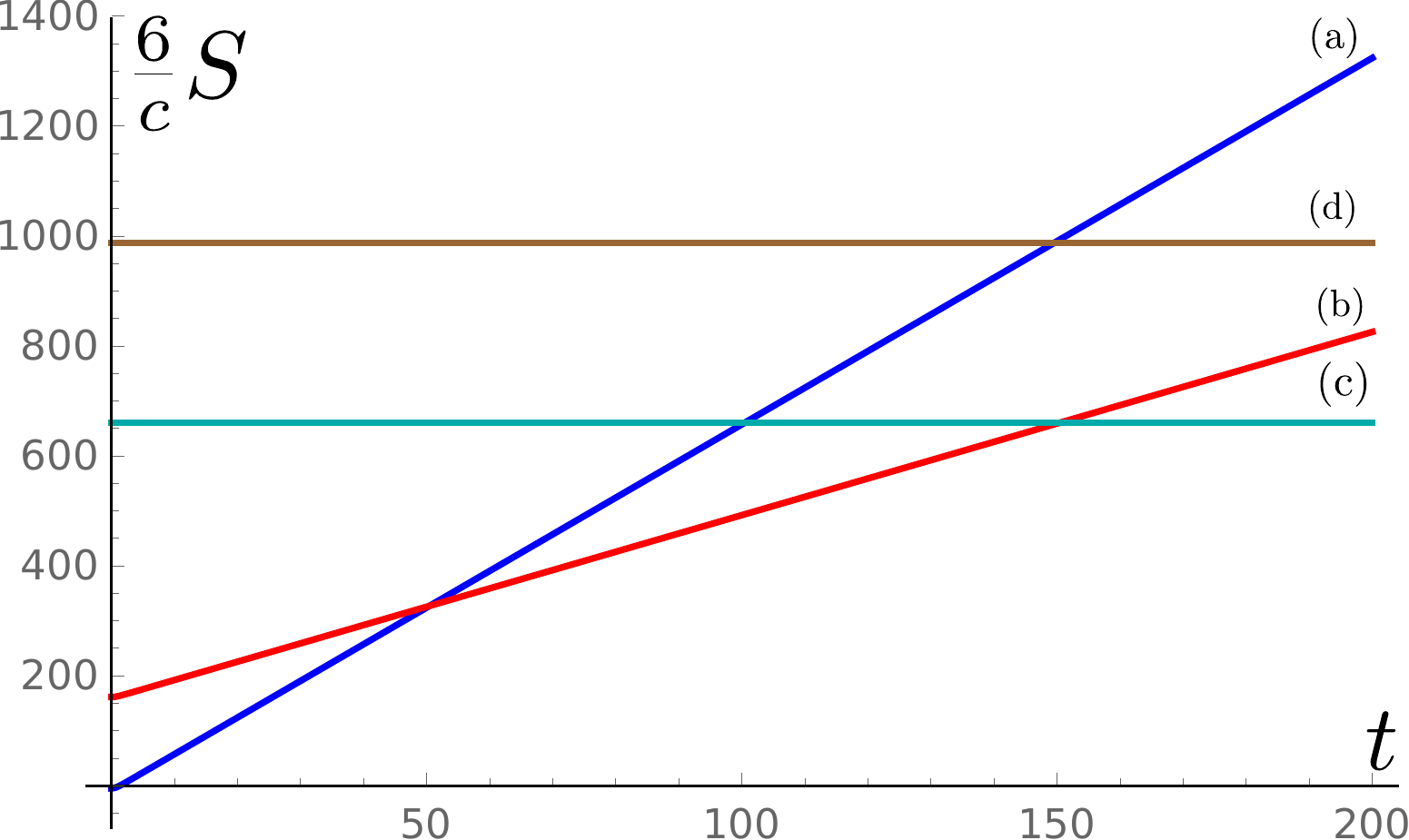}
\includegraphics[scale=0.43]{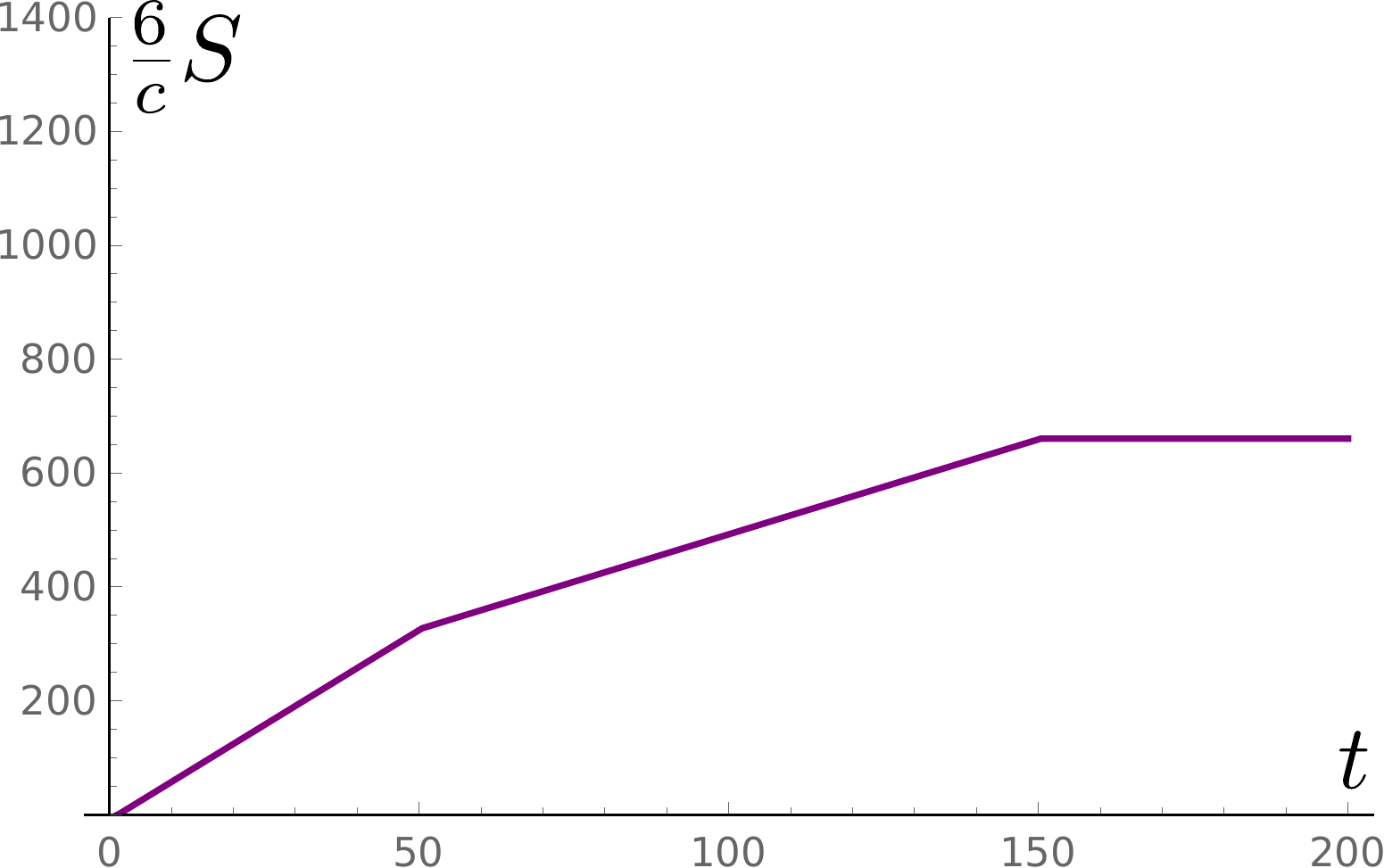}\\
\textbf{A.} \hspace{6cm} \textbf{B.}
\caption{\textbf{A.} The entanglement entropy curves for the configurations (a), (b), (c) and (d) for region size much larger than the distance from the interface and comparable to $\phi_r/c$. \textbf{B.} The Page curve of a finite segment, obtained by minimizing between the saddles shown in plot A.}
\label{fig:PageCurve1BH}
\end{center}
\end{figure}

\paragraph{Summary: }Varying the size of the reservoir regions with other parameters fixed, we see in Fig.~\ref{fig:PageCurve1BH} that for  large regions we have  transitions (a) $\rightarrow$ (b) $\rightarrow$ (c), accessing an island region for a finite period of time before losing it. For smaller regions we directly make the thermalization transition (a) $\rightarrow$ (c) as in \cite{HM}.  Thus, finite radiation segments give temporary access to the black hole interior, unlike infinite segments which give permanent access at late times \cite{Almheiri193,Almheiri194,Penington192,Almheiri195}.  This happens because  during the entanglement evolution any finite segment will  eventually thermalize,  scrambling  information from the island.  From the path integral point of view the transition (a) $\rightarrow$ (b) arises by including replica wormholes \cite{Almheiri193, Penington192}, and will be smoothed by also summing over replica non-symmetric manifolds \cite{Penington192, Dong-Marolf,Dong-Marolf1}.

By utilizing the unitarity bound discussed in section~\ref{intro}, the qualitative nature of these transitions can be predicted without much computation. In particular, (a) and (b) will eventually threaten the unitarity bound set by the size of the Hilbert space of our radiation region, so must eventually transition. What about the transition (a) $\rightarrow$ (b)? Clearly this need not come close to saturating the unitarity bound of our radiation region, since the entropy continues to increase in phase (b) due to interior Hawking modes being captured by the island while their exterior partners escape away into the infinite region of the bath. However, the structure of the geodesics in Fig.~\ref{fig:1BH-geod}(a) indicates that the mutual information vanishes between (i) the gravitational region and the adjoining baths up to the radiation region, and (ii) the rest of the complement of the radiation region $A$. In particular, this means that each region of the complement is subject to its own unitarity bound. The relevant bound here is the one set by the gravitational region (plus some of the adjoining baths, which we take to be a small correction): once the entropy of this region comes close to $2S_{\text{BH}}$, a transition must occur.

%%%%%%%%%%%%%%%%%%%%%%%%%%%%%%%%%%%%%%%%%%%%%%%%%%%%%%%%%%%%%%%%%%%%%%%%%%%%%%%%%%%%%%%%%%%%%%%%%
\section{Two holographic quantum dots connected by a reservoir}\label{sec3}

The quantum-mechanical setup we consider next involves two thermofield double pairs of a holographic quantum mechanical system (which we refer to as the quantum dot) interacting through a common reservoir. The quantum dots are dual to  JT gravity described by the action (\ref{JTact}). They are coupled to a CFT$_2$ of central charge $c$, which lives in the reservoir. We will not specify the exact Hamiltonian of the quantum dots; however a common example of a quantum mechanical system holographically dual to JT gravity (in an appropriate limit) is the SYK model \cite{Sachdev92,Kitaev}.

\begin{figure}[t]
	\centering
	\includegraphics[scale=0.5]{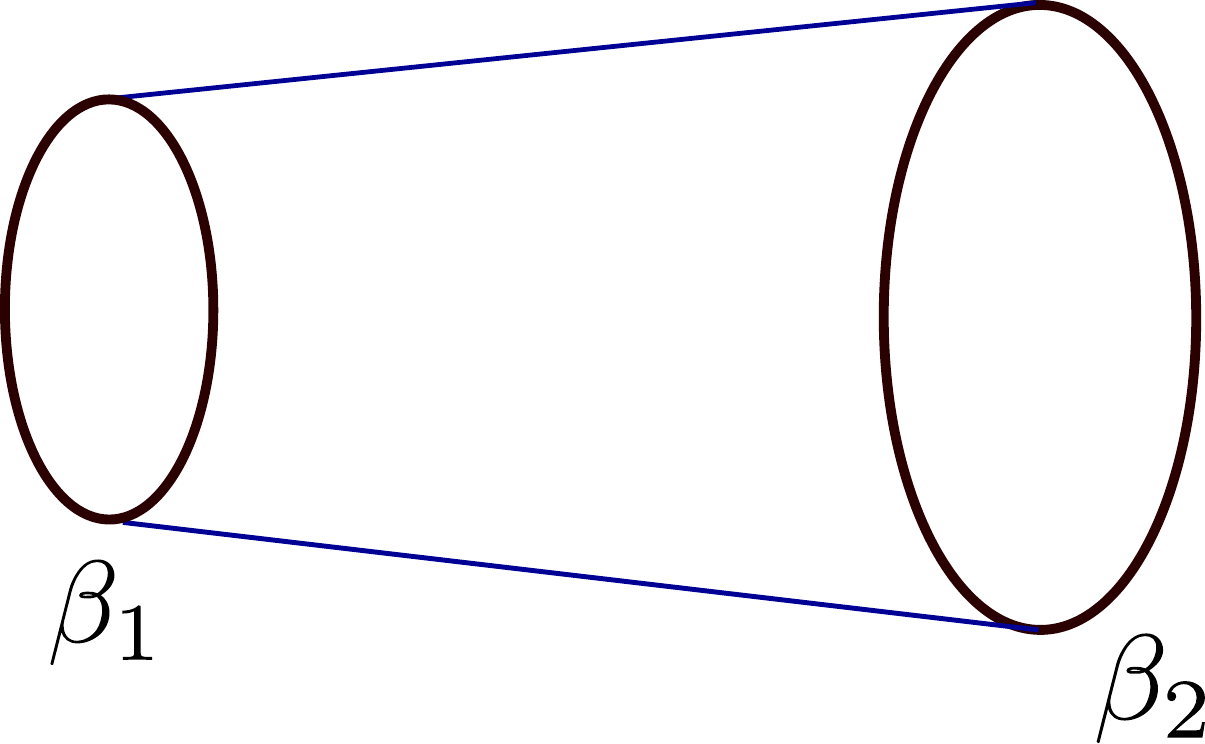}
	\caption{Euclidean picture of the two quantum dots living on the thermal circles of different lengths coupled by the conical reservoir. }
	\label{fig:SYK-bCFT-SYK}
\end{figure}

The temperatures of the quantum dots are treated as independent parameters, and  so for this setup to be in the equilibrium, we have the reservoir working as a heat engine which cools the matter CFT quanta being emitted from the hotter quantum dot and going into the cooler one. The action of the heat engine is caused by the nontrivial metric in the reservoir. Such a reservoir in Euclidean signature has a boundary consisting of two thermal circles of different lengths (Fig.~\ref{fig:SYK-bCFT-SYK}). One can write down a metric for such a reservoir  as follows:\footnote{We include the regulator $\epsilon$ so that the metric in the gravity dual is manifestly continuous throughout the spacetime, similarly to the discussion in Sec.~\ref{sec:1xBH}.}
\be
ds^2 = \frac{dr^2 + f(r)^2 d\tau^2}{\epsilon^2}\,, \qquad \tau \sim \tau  + 2\pi\alpha\,,\qquad r \in (r_1, r_2)\,.
\ee
Notice that the thermal periodicity changes from $2\pi f(r_1)\alpha/\epsilon$ to $2\pi f(r_2)\alpha / \epsilon$. So, as advertised, such a spacetime acts as a heat engine which equilibrates the radiation between the two systems with different temperature. This is similar to considering different radial positions in a black hole in thermal equilibrium: the physical temperature for static observers at each location is different, yet the entire system is in thermal equilibrium.

In the present paper we consider the simplest case and assume that $f(r) = r$ and the reservoir has the metric of a cone:
\be
ds_R^2 = \frac{dr^2 +  r^2 d\tau^2}{\epsilon^2}\,,\qquad \t \sim \t + 2\pi \a \label{Rindler}\,.
\ee
The results in this model can be extended to a general $f(r)$ in a straightforward manner. The position-dependent temperature is precisely that of Rindler space, obtained by continuing to Lorentzian signature $\t \rightarrow it$.

By choosing an interval of size $L_R$ and inverse temperatures $\b_1$ and $\b_2$, we fix the radial coordinates of boundaries in Fig.~\ref{fig:SYK-bCFT-SYK} and the solid angle of the cone. Choosing $\beta_2 > \beta_1$, this yields the relations
\be 
r_1 = \frac{L_R\beta_1}{\b_2 - \b_1}\,, \qquad r_2 = \frac{L_R\beta_2}{\b_2-\b_1}\,,\qquad \a = \f{\b_2-\b_1}{2\pi L_R}\,. \label{r12}
\ee
For $\a \neq 1$ there is a conical singularity, although it is excised from the region we are considering. The parameter $\a$ has a physical interpretation as the strength of the heat engine which equilibrates the two sides, and it must be tuned as we vary the temperatures or the distance over which we want to equilibrate the temperatures. Some formulas will be simpler to write in terms of $\alpha$, but  fixing $\alpha$ and $\b_{1,2}$ fixes $L_R$. The equal-temperature limit can be realized by taking $\alpha\to 0$ while keeping $L_R$ fixed, in which case $r_{1,2} \rightarrow \infty$. Physically we are taking the strength of the equilibrator to zero (equal temperatures are already equilibrated), but to keep finite-sized circles we need to scale the coordinates to infinity. 

We can rewrite the cone metric as conformally flat: 
\be\label{MetricRi}
ds_R^2 = \Omega_R(\xi)^{-2} \frac{d \xi^2 + d \theta^2}{\epsilon^2}\,, \qquad \xi \in \left[0, L\right]\,,\qquad \theta \sim \theta + 2\pi\,.
\ee
Note that the dimensionless coordinate $\xi$ plays the role of the spatial coordinate, and $\theta$ plays the role of Euclidean time. The Weyl factor is given by the formula  
\be \label{OmegaR}
\Omega_R (\xi) = \frac{1}{\a r} = \frac{2\pi}{\beta_1} e^{-\alpha \xi}\,,\qquad r =r_1 e^{\a \xi}\,,\qquad L = \a^{-1}\log\f{\b_2}{\b_1}\,.
\ee
\begin{comment}
Then the Weyl factor reads 
\be
\Omega_R = \frac{\beta}{\beta_1} \left(\frac{\beta_1}{\beta_2}\right)^{\frac{\sigma}{L}}\,. \label{OmegaR}
\ee
Notice that if we want to keep $L$ fixed as we vary the temperatures then we need to vary $\g$.  In particular this means we use \emph{different} geometries as we vary $L$ (in particular the conical angle changes). This is reasonable, since fixing $L$, $\b_1$, and $\b_2$ fixes both the conical geometry and which piece of it we cut out. More physically, as we make the temperatures very different we need a stronger heat/engine to equilibrate them.
\end{comment}
There are two possible two-dimensional gravitational duals for our Rindler reservoir coupled to holographic quantum dots living on thermal circles at each end. We can either complete the circular boundaries of the reservoir into a torus, or we can have independent disks at the two boundaries. The first possibility represents a confined phase, described by a wormhole, shown schematically in Fig.~\ref{fig:WH}. The second possibility represents a deconfined phase, described by independent thermofield double black holes, shown schematically in  Fig.~\ref{fig:SYK-bCFT-SYK}.  

The partition sum of the theory will undergo transitions between these confined and deconfined phases as the parameters change.  We expect that the phase structure will be  similar to the one in \cite{Chen20}. The wormhole solution should be global AdS$_2$ as in \cite{Chen20}, except that the dilaton profile will be a one-parameter generalization of the usual one: the solution has a free constant that is usually fixed by the $\mathbb{Z}_2$ symmetry, but in our case this constant will be fixed by the choice of  temperatures. This allows us to cut the solution off at slightly different coordinate values on either side of the wormhole, which leads to equal values of $\phi_r$, but with different temperatures. In other words, the boundary conditions from the bulk perspective are given by 
\be
ds^2_{1,2} = \f{du^2}{\epsilon^2}\,,\quad u \sim u + \b_{1,2}\,,\qquad \phi_{1,2} = \f{\phi_r}{\epsilon}
\label{phirdef}
\ee
at leading order in $\epsilon$. 
The other  novelty is the Rindler region. As we reviewed above, a Weyl transformation maps this to the usual case of a finite cylinder with constant thermal periodicity. The Weyl anomaly then makes a contribution to the partition sum, but since the transformation is the same in both solutions, the anomaly  does not affect the phase structure. 

\subsection{Black hole phase}
\label{sec:BHphase}

The action is given in \eqref{JTact} and the solution that describes the geometry (Fig.~\ref{fig:SYK-bCFT-SYK}) involves two Euclidean copies of the gravitational solution in \eqref{BHmetsol}-\eqref{BHdilsol}, with different temperatures:
\bea
ds_1^2 &=& \frac{4\pi^2}{\beta_1^2} \frac{d\tau_1^2+d\sigma^2}{\sinh^2 \frac{2\pi \sigma}{\beta_1} }\,; \quad \sigma \in (-\infty, -\epsilon]\,,\qquad \tau_1 \sim \tau_1 + \b_1 \,,\label{MetricPhysE1b1}\\
ds_2^2 &=& \frac{4\pi^2}{\beta_2^2} \frac{d\tau_2^2+d\sigma^2}{\sinh^2 \left[\frac{2\pi\sigma}{\beta_2} - L\right]}\,;\quad \sigma \in \left[\frac{\beta_2}{2\pi} L + \epsilon, +\infty\right)\,,\qquad \tau_2 \sim \tau_2 + \b_2\,. \label{MetricPhysE2b2}
\eea
The reservoir metric is given by (\ref{MetricRi}), with $L = \alpha^{-1} \log \frac{\beta_2}{\beta_1}$ being the size of the reservoir in the cylinder coordinates. The reservoir coordinate $\xi$ is dimensionless, hence the extra conversion factor of $\frac{\beta_2}{2\pi}$. The dilaton profile in the disk regions is given by the equations 
\bea
\phi(\sigma)_{1} &=& -\frac{2\pi}{\beta_1} \phi_r \coth \frac{2\pi \sigma}{\beta_1}\,; \label{dilaton-BH-1-b1}\\
\phi(\sigma)_{2} &=& \frac{2\pi}{\beta_2} \phi_r \coth \left(\frac{2\pi \sigma}{\beta_2} - L\right)\,. \label{dilaton-BH-2-b2}
\eea

We can break up the on-shell partition function for this solution into two main pieces. The first  is the gravitational contribution on the disks; it is obtained by directly evaluating the pure JT gravity part of the action on the solution, and reads 
\be
Z_{\text{disks,grav}} = e^{2\phi_0 + \pi \phi_r \left(\frac{1}{\beta_1} + \frac{1}{\beta_2}\right)}\,.  \label{Z-BHgrav}
\ee
The second piece of the on-shell partition function comes from the matter CFT living on the curved background. The spacetime metric can be written as $ds^2 = e^{2\omega} d\hat{s}^2_{\text{flat}}$ everywhere in the spacetime, and so the nontrivial Weyl factor $\omega$ gives  a Weyl anomaly contribution \cite{Polyakov}:
\be\label{lianom}
Z_{\text{anomaly}} = \exp\left[\f{c}{24\pi} \int \sqrt{\hat{g}} \,(\partial \omega)^2\right].
\ee
The computation of this partition function involves a step which will also be used in the computations of entanglement entropy in this model, so we discuss it in detail. 

Our spacetime is a closed manifold, so there are no physical boundaries to generate boundary terms. We perform the conformal transformation from the disks and the conical reservoir shown in Fig.~\ref{fig:SYK-bCFT-SYK} to the plane with metric of the form $ds^2 = e^{2\omega} d\hat{s}^2_{\text{flat}}$. We write the flat reference metric in  polar coordinates: 
\be
d\hat{s}^2_{\text{flat}} = d w d \bar{w} = d\rho^2 + \rho^2 d\varphi^2\,,
\ee 
where $w = \rho e^{i \varphi}$ is the complex coordinate, $\rho \in [0, +\infty)$ and $\varphi \sim \varphi + 2\pi$. The explicit form of the conformal transformation is given by  
\bea
\text{Disk 1}: &&\ w = e^{\frac{2\pi}{\beta_1} (\sigma + i\tau_1)}\,; \\
\text{Reservoir}: &&\ w = e^{\xi + i\theta}\,; \\
\text{Disk 2}: &&\ w = e^{\frac{2\pi}{\beta_2} (\sigma + i\tau_2)}\,.
\eea
\begin{figure}[t]
	\centering
	\includegraphics[scale=0.35]{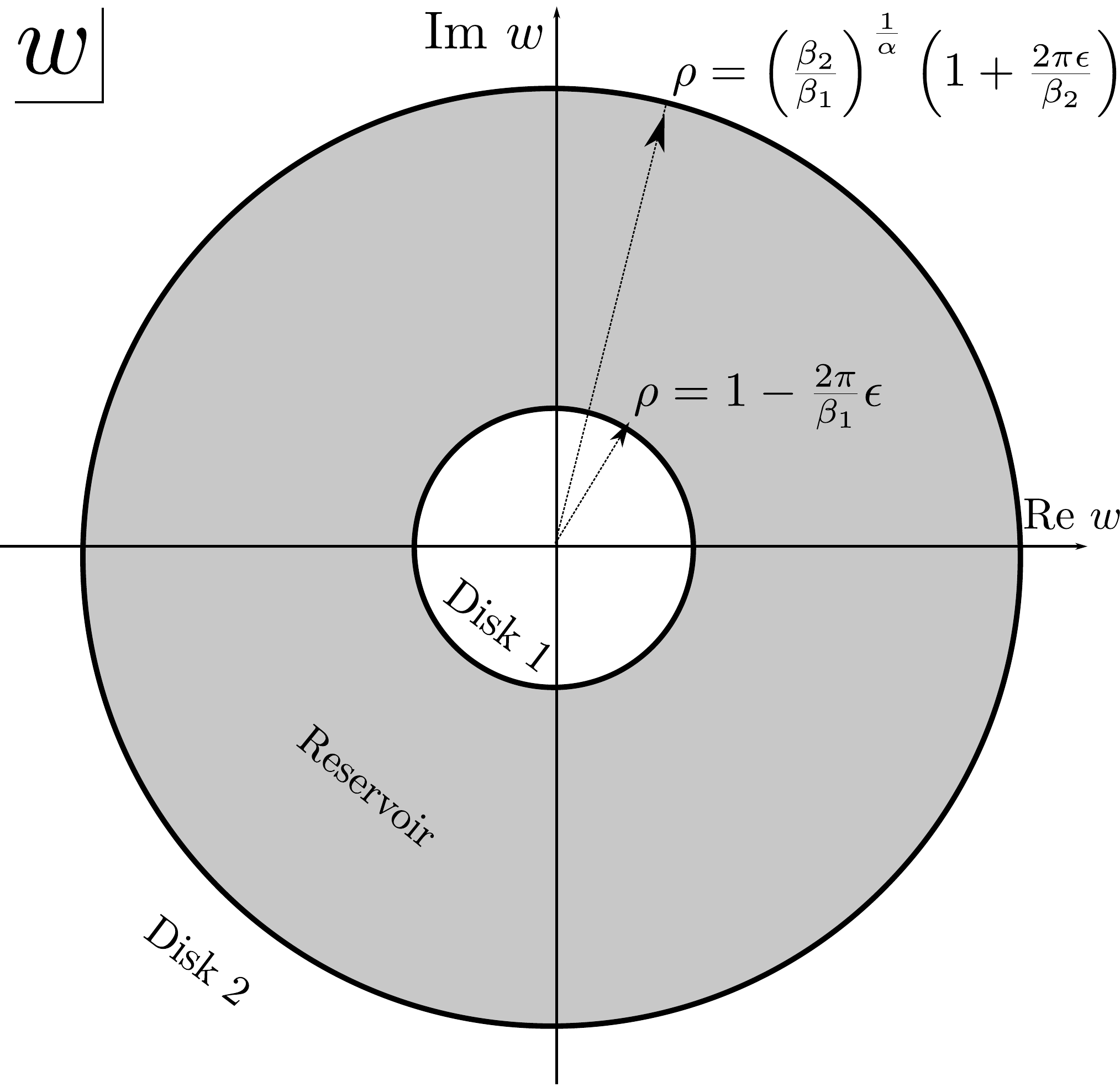}  
	\caption{Euclidean geometry of two black holes coupled via a reservoir conformally mapped to a plane. The shaded region is the reservoir.}
	\label{fig:SYK-bCFT-SYK-plane}
\end{figure}
As shown in Fig.~\ref{fig:SYK-bCFT-SYK-plane}, the plane is divided into three regions where the Weyl factor $\omega$ is given by different functions: 
\bea
\text{Disk 1}: && e^{\omega_1} = \Omega_1^{-1} = \frac{2}{1-\rho^2}\,, \qquad \rho \in \left[0, 1-\frac{2\pi}{\beta_1}\epsilon\right]\,;\label{Omega1}\\
\text{Reservoir}: && e^{\omega_{\text{bath}}} =\Omega_{\text{bath}}^{-1} = %\left(\epsilon \Omega_R \times \rho\right)^{-1} = 
\frac{\beta_1}{2\pi\epsilon} \rho^{\alpha-1}\,, \ \  \rho \in \left(1-\frac{2\pi}{\beta_1}\epsilon,  \left(\frac{\beta_2}{\beta_1}\right)^{\frac{1}{\alpha}} \left(1 + \frac{2\pi \epsilon}{\beta_2}\right) \right); \label{Omegabath}\\
\text{Disk 2}: && e^{\omega_2} = \Omega_2^{-1} = \frac{2 \left(\frac{\beta_2}{\beta_1}\right)^{\frac{1}{\alpha}}}{\rho^2 - \left(\frac{\beta_2}{\beta_1}\right)^{\frac{2}{\alpha}}}\,,  \quad \rho \in \left[\left(\frac{\beta_2}{\beta_1}\right)^{\frac{1}{\alpha}} \left(1 + \frac{2\pi \epsilon}{\beta_2}\right), \left(\f{\b_2}{\b_1}\right)^{\f{1}{\alpha}} \Lambda\right]\,,\label{Omega2}
\eea
where the bottom line includes the Weyl factor from the Rindler geometry in the reservoir defined in (\ref{OmegaR}) written in terms of $w = e^{\xi+ i\theta}$, and $\Lambda \gg 1$ is the cutoff. The cutoff is defined with the $(\b_2/\b_1)^{1/\alpha}$ prefactor to ensure the ratio of radii defining disk 2 is independent of $\b_1$, which means the contribution of disk 2 to the partition function will be independent of $\b_1$ (see \eqref{Z2} below). We now compute the contributions from the three regions to (\ref{lianom}) separately. 

\paragraph{Disk 1:} We integrate from $\rho = 0 $ to $\rho = 1-\frac{2\pi}{\beta_1} \epsilon$ and get
	\be
Z_1 = \exp\left[ \frac{c \beta_1}{24\pi \epsilon}+\f c 6 \log \frac{4\pi\epsilon}{\beta_1} -\f c 8 \right]\tilde{Z}_{1}\,, \label{Z1}
\ee 
where $\tilde{Z}_1$ is the partition function of the theory on the portion $\rho \in [0,1-2\pi \epsilon/\b_1]$ of flat space. 

\paragraph{Disk 2:} We integrate from $\rho = \left(\frac{\beta_2}{\beta_1}\right)^{\frac{1}{\alpha}} \left(1 + \frac{2\pi \epsilon}{\beta_2}\right) $ to $\rho = \Lambda \gg 1$ and get
\be
Z_2 = \exp\left[\frac{c \beta_2}{24\pi \epsilon}-\f c 6 \log \frac{4\pi\epsilon}{\beta_2} -\f{c}{24}+\f c 3 \log \L \right]\tilde{Z}_2\,, \label{Z2}
\ee
where $\tilde{Z}_2$ is the partition function of the theory on the portion $\rho \in \left[\left(\frac{\beta_2}{\beta_1}\right)^{\frac{1}{\alpha}} \left(1 + \frac{2\pi \epsilon}{\beta_2}\right), \left(\frac{\beta_2}{\beta_1}\right)^{\frac{1}{\alpha}}\Lambda\right]$ of flat space. 

\paragraph{Reservoir region:} Here we integrate from $\rho = 1-\frac{2\pi}{\beta_1}\epsilon$ to  $\rho = \left(\frac{\beta_2}{\beta_1}\right)^{\frac{1}{\alpha}} \left(1 + \frac{2\pi}{\beta_2}\epsilon\right) $ and get
\be
Z_{\text{reservoir}} = \exp\left[\frac{c (\alpha-1)^2}{12 \alpha} \log \frac{\beta_2}{\beta_1}\right]\tilde{Z}_{\text{reservoir}}\,, \label{Zres}
\ee
where $\tilde{Z}_{\text{reservoir}}$ is the partition function of the theory on the portion $\rho \in [1-2\pi \epsilon/\b_1, \left(\frac{\beta_2}{\beta_1}\right)^{\frac{1}{\alpha}} (1 + 2\pi\epsilon/\b_2)]$ of flat space. Combining (\ref{Z1}) - (\ref{Zres}) and using the normalization $\tilde{Z}_1 \tilde{Z}_2 \tilde{Z}_{\text{reservoir}} = 1$, we get the total result for the CFT anomaly contribution: 
\be
Z_{\text{anomaly}} = \exp\left[-\frac{c}{6} + \frac{c }{12} \left(\alpha + \frac{1}{\alpha}\right) \log \frac{\beta_2}{\beta_1} + \frac{c (\beta_1 + \beta_2)}{24 \epsilon} +\f c 3 \log \L \right] . \label{Z-BHan}
\ee
Next we combine this result with the gravitational partition function (\ref{Z-BHgrav}):
\be
Z = \exp\left(2\phi_0 -\f{c}{6}+ \pi \phi_r\left(\f{1}{\b_1}+\f{1}{\b_2}\right)+\frac{c }{12} \left(\alpha + \frac{1}{\alpha}\right) \log \frac{\beta_2}{\beta_1}\right)\,. \label{Z-BH}
\ee
This is the final result for the partition function of the disconnected black hole phase. We removed two terms from (\ref{Z-BHan}): a  $1/\epsilon$ divergence proportional to the sum of lengths of the boundary of the hyperbolic disk which was removed by a local counterterm, and a divergence in $\L$ that is expected to renormalize $\phi_0$.\footnote{Note  that if we chose the cutoff in the plane as $\L$ instead of $(\b_2/\b_1)^{1/\alpha} \L$, it would modify the coefficient of the term $\a^{-1} \log \f{\b_2}{\b_1}$ and give the wrong answer in the equal-temperature limit where $\a \rightarrow 0$ and $\a^{-1}\log \f{\b_2}{\b_1} \rightarrow 2\pi L_R/\b$, where we denote the limiting value of $\b_1$ and $\b_2$ as $\b$.} 
%%%%%%%
\subsection{Asymmetric wormhole phase}
\label{sec:WHphase}

\begin{figure}[t]
	\centering
		\includegraphics[scale=0.3]{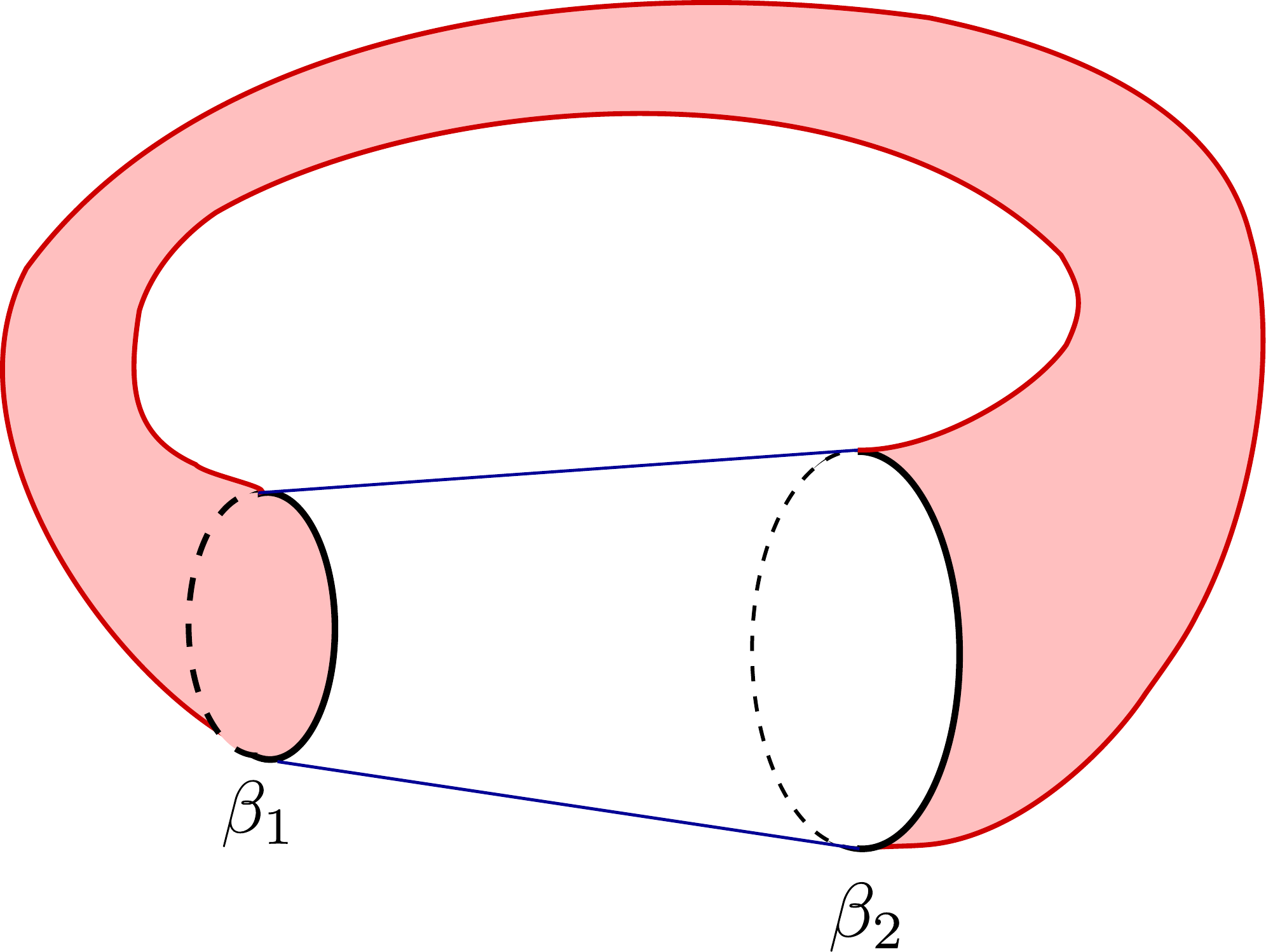}
	\caption{Schematic  of the wormhole phase. The asymmetry of the wormhole is caused by the different sizes of the thermal circles which are glued to the regulated AdS$_2$ boundaries. Correspondingly, the two boundary regulators are different. }
	\label{fig:WH}
\end{figure}

The asymmetric wormhole solution  is a quotient of Euclidean AdS$_2$ spacetime with the metric  \cite{Maldacena18,Chen20}
\be
ds^2 = \frac{d\chi^2 + d\sigma^2}{\sin^2 \sigma}\,,\qquad \phi = -2\pi T_{\s\s}^{\text{traceless; bulk}} \left(\f{\g-\s}{\tan \s}+1\right)\,, \label{AdS2}
\ee
where $\chi \sim \chi + b$ is a periodic coordinate which plays the role of Euclidean time direction and $\sigma \in [\epsilon_1, \pi - \epsilon_2]$ is the spatial coordinate with $\epsilon_{1,2}$ being small cutoffs to be determined below. The two boundaries of AdS$_2$ are glued to the Rindler reservoir with the metric (\ref{MetricRi}). The spacetime is thus a conformally flat manifold with the topology of a torus, as shown in Fig. \ref{fig:WH}. 

It remains to compute $T_{\s\s}^{\text{traceless; bulk}}$ and fix $\g$. Ignoring the curved metric for a moment, the torus has circumferences $b$ and $d = \pi+ (2\pi \a)^{-1}b \log \f{\b_2}{\b_1}$ along the $\chi$ and $\sigma$ directions of the AdS$_2$ bulk, and along the $\theta$ and $\xi$ directions of the reservoir,  respectively. For the wormhole solution to exist, we take $b>d$.  If we consider a holographic CFT on this background, and quantize along the $\chi$ direction of the AdS$_2$ ($\theta$ direction of the reservoir),  we are in the vacuum.   The vacuum stress tensor on such a cylinder of circumference $d$, evaluated for a torus with the flat metric $\hat{g}_{\mu\nu}$ is therefore: 
\be
T_{\s\s}^{\hat{g}} = - T_{\c\c}^{\hat{g}} = -\f{\pi c}{6d^2}= -\f{c}{6\pi} \f{1}{\left(1+\f{b \log \f{\b_2}{\b_1} }{2\pi^2 \a}\right)^2}\,.
\ee
Now recall that the gravitational part of the manifold has a curved metric \eqref{AdS2} of the form $g_{\mu\nu} = e^{2 \omega} \hat{g}_{\mu\nu}$, where $\hat{g}_{\mu\nu}$ is the flat metric. This gives an anomalous contribution to the stress tensor which has a traceless piece and a piece proportional to the metric. The traceless piece coming from the combination of the flat torus and the anomaly is what enters in the dilaton solution above and is given as 
\be
T_{\s\s}^{\text{traceless; bulk}} = - T_{\c\c}^{\text{traceless; bulk}} = -\f{c}{6\pi} \f{1}{\left(1+\f{b \log \f{\b_2}{\b_1} }{2\pi^2 \a}\right)^2} + \f{c}{24\pi}\,.
\ee
Unlike previous works we have a free parameter $\g$ in the solution for the dilaton; without this free parameter the solution would only exist for equal temperatures on the two sides. We want to match the lengths of the different thermal circles at the interfaces between the wormhole and the Rindler bath while maintaining equal $\phi_r$ values (see Eq.~\ref{phirdef}). This requires asymmetric cutoffs $\s_1=\epsilon_1\ll 1$, $ \pi-\s_2 =\epsilon_2 \ll 1$: 
\be
\frac{b}{\epsilon_{1,2}} = \frac{\beta_{1,2}}{\epsilon} \implies  \epsilon_{1,2} = \frac{b}{\beta_{1,2}}\,\epsilon\,,\qquad \phi\big|_{\s_1}=  \phi\big|_{\s_2} \implies \g = \f{\pi \b_2}{\b_1+\b_2}\,. \label{cutoffs-wormhole}
\ee
So the choice of temperatures $\b_1$, $\b_2$ dictates the cutoffs $\s_1$, $\s_2$ and the dilaton solution. In the symmetric case $\b_1 = \b_2$ we would find $\g = \pi/2$. The dilaton boundary condition fixes $b$ through
\be\label{bsol}
\phi\big|_{\s_1} = \f{\phi_r}{\epsilon} \implies  \left(\frac{c}{3\left(1+\frac{b \log \f{\b_2}{\b_1}}{2\pi^2\a }\right)^2} - \frac{c}{12}\right)= \f{\phi_r b(\b_1+\b_2)}{\pi \b_1 \b_2}\,.
\ee

Having established the dilaton solution, we are now ready to discuss the full partition function of the wormhole phase. We can break up the partition function in this phase into three pieces: 
\begin{itemize}
    \item[(i)] The gravitational contribution from the AdS$_2$ wormhole, computed using the action (\ref{JTact}) evaluated on the solution (\ref{AdS2}). 
    \item[(ii)] The  CFT anomaly from the Weyl transform from cylinder $\chi \in [0, b)$, $\sigma \in [\epsilon_1, \pi - \epsilon_2]$ to the AdS$_2$ metric in the gravitational region. It is computed using equation (\ref{lianom}) with $e^{\omega} = \frac{1}{\sin \sigma}$. 
    \item[(iiia)] The contribution of the CFT on a torus of lengths $d$ and $b$. Recall that we assume $b > d$ to project onto the vacuum for a holographic CFT. Under this assumption the partition function is the thermal partition function of a CFT on the circle of size $d = \pi+ (2\pi \a)^{-1}b \log \f{\b_2}{\b_1}$. 
    \item[(iiib)] The CFT anomaly from Weyl transforming from the cylinder to the cone in the reservoir region with the Weyl factor given by (\ref{OmegaR}).
\end{itemize}
Altogether we get 
\be
Z = Z_{(i)}Z_{(ii)}Z_{(iii)} = \exp\left(\underbrace{-\f{\phi_r b^2}{4\pi}\left(\f{1}{\b_1}+\f{1}{\b_2}\right)}_{(i)}\,\,\underbrace{-\,\,\f{cb}{24}}_{(ii)}+\underbrace{\f{c}{6}\f{b}{1+\f{ b\log\f{\b_2}{\b_1}}{2\pi^2 \a}}}_{(iiia)}+\underbrace{\f{c \a}{12} \log\f{\b_2}{\b_1}}_{(iiib)}\right) \label{Z-WH}\,.
\ee
Note that the contribution (i) is accompanied by a $1/\epsilon$-divergence proportional to the length of AdS boundaries, which is identical to the $1/\epsilon$-divergence that appeared in (\ref{Z-BHan}) in the disconnected phase and is removed in the same way. Note that the equation (\ref{bsol}) can be reproduced by extremizing (\ref{Z-WH}) over $b$.

The contribution (iiib) is the same between the two phases, and can hence be ignored in determining which phase is dominant. The phases are easy to compare in two limits for which the equation (\ref{bsol}) has a simple solution. In the first limit we have (trading $\a$ for $L_R$)
\begin{align}
L_R \ll \phi_r/c :\qquad &b \approx \f{\b_1 \b_2 c \pi}{4 \phi_r (\b_1 + \b_2)}\\
 \implies\, Z_{WH} \approx \exp\left(\f{\b_1 \b_2 c^2 \pi}{64 \phi_r (\b_1 +\b_2)} + I_R\right),\,\qquad  &Z_{BH} \approx \exp\left(2\phi_0 + \left(\f{1}{\b_1} + \f{1}{\b_2}\right)\pi \phi_r + I_R\right)\,,
\end{align}
where $I_R = \frac{c}{24 \pi}\frac{\beta_2 - \beta_1}{L_R} \log \frac{\beta_2}{\beta_1}$ is the contribution from the reservoir which is common for the two phases. Equating the two expressions, we find that the wormhole dominates for
\be
\f{\b_1 \b_2}{\b_1 + \b_2} \gtrsim \f{128 \phi_r \phi_0}{c^2 \pi}\,.
\ee
In the second limit we have
\begin{align}
L_R \gg \phi_r/c :\qquad &b \approx \f{\pi(\b_2-\b_1)}{L_R\log \f{\b_2}{\b_1}}\\
 \implies\, Z_{WH} \approx \exp\left(\f{c\pi (\b_2-\b_1)}{24L_R\log \f{\b_2}{\b_1}} + I_R\right),\,\qquad  &Z_{BH} \approx \exp\left(2\phi_0 - \f{\pi c L_R \log \f{\b_2}{\b_1}}{6(\b_2-\b_1)}+ I_R\right)\,.
\end{align}
Equating the two expressions, we find that the wormhole dominates for
\be
\f{\b_2-\b_1}{\log \f{\b_2}{\b_1}} \gtrsim \f{48 L_R \phi_0}{c \pi}\,.
\ee
Notice that in this limit it is sufficient to take $\b_2$ large at any value of $\b_1$ for the wormhole to dominate. 

%%%%%%%%%%%%%%%%%%%%%%%%%%%%%%%%%%%%%%%%%%%%%%%%%%%%%%%%%%%%%%%%%%%%%%%%%%%%%%%%%%%%%%
\section{Page curves of two connected black holes} \label{sec4}

We can now analyze the structure of entanglement entropy in our model to ask when island configurations appear, and whether and how they compete with other quantum extremal surfaces to saturate the entropy. 

A priori, we can study this question in either of the two phases discussed in the previous section. However, we will argue that the wormhole phase does not have any island saddles. First we show that, for a holographic theory, we have to be in the vacuum channel, i.e.\ in the doubly holographic picture the three-dimensional bulk filling in the torus is thermal AdS$_3$. To see this, if we were instead in the vacuum in the dual channel, i.e.\ the three-dimensional bulk filling in the torus being the BTZ black hole, this would imply a positive stress-energy tensor. Such a stress tensor would not consistently solve the equation of motion (\ref{dilatonEOM}) with the boundary conditions needed for the dilaton. This is the familiar statement that we need a negative stress-energy to support the wormhole \cite{GJW,Maldacena18}.\footnote{The spacetime wormhole can be converted into an (eternally) traversable spatial wormhole by analytic continuation, e.g. in equation \eqref{AdS2} continue $\chi \to it$.} To complete the argument, note that in the vacuum channel the 2d entanglement entropy is $O(c)$ and constant for intervals in the bath and their thermofield double partners. Any nontrivial QES will come with a cost of $\phi_0$, which we assume to be much larger than the CFT central charge $c$. Hence the nontrivial QES will never dominate. In the rest of this section we therefore focus on the phase with two black holes.

%%%%%%%
\subsection{The two black hole setup}
\begin{figure}[t]
	\centering
	\includegraphics[scale=0.8]{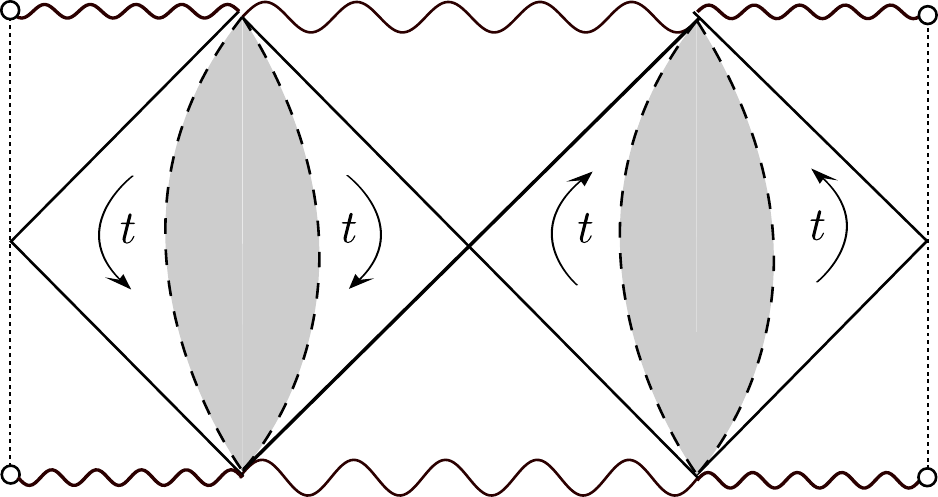} 
	\caption{The Penrose diagram for two eternal black holes with different inverse temperatures $\beta_1$, $\beta_2$ connected by two copies of the reservoir (shown by the shaded regions) with the conical metric (\ref{MetricRi}). The arrows show the direction of the coordinate time in the external regions of the black holes. The diagram is identified across the dashed line which goes through the bifurcation surface of the second black hole.}
	\label{fig:EE-2sided}
\end{figure}
We are interested in the temporal behavior of the entanglement entropy of the two thermofield double copies of a segment that includes some portion of radiation in the reservoir and may or may not include one of the quantum dots dual to black holes. The metric of the exterior regions of the black hole is given by the Lorentzian version of the solution  (\ref{MetricPhysE1b1})-(\ref{MetricPhysE2b2}). We rescale the coordinates in such a way that they are dimensionless and continuous across the pairs of exterior regions connected to their corresponding reservoirs. The metric in the external black hole region then reads: 
\bea
ds_1^2 &=& \frac{-dt^2+d\xi^2}{\sinh^2 \xi }\,; \quad \xi \in \left(-\infty, - \frac{2\pi}{\beta_1}\epsilon\right] \label{MetricPhysL1b1}\\
ds_2^2 &=& \frac{-dt^2+d\xi^2}{\sinh^2 (\xi - L)}\,;\quad \xi \in \left[L + \frac{2\pi}{\beta_2}\epsilon, +\infty\right) \label{MetricPhysL2b2}\,, 
\eea
and the reservoir metric is the Lorentzian continuation of the metric (\ref{MetricRi}): 
\be
ds_{R}^2 = \frac{\beta_1^2}{4\pi^2} e^{\alpha \xi} \frac{-d t^2 + d\xi^2}{\epsilon^2}\,, \quad \xi \in \left[- \frac{2\pi}{\beta_1}\epsilon, L+\frac{2\pi}{\beta_2}\epsilon\right]\,. \label{MetricRiL}
\ee
Note that temperature dependence is now contained in the cutoffs for $\xi$. Thus the lengths of thermal circles on the boundaries in the Euclidean continuation of the metric  are the same and are equal to $\beta_{1,2}/\epsilon$. The dilaton solution in these rescaled coordinates reads 
\bea
\phi(\sigma)_{1} &=& -\frac{2\pi}{\beta_1} \phi_r \coth (-\xi)\,; \label{dilaton-BH-1-b1r}\\
\phi(\sigma)_{2} &=& \frac{2\pi}{\beta_2} \phi_r \coth \left(\xi - L\right)\,. \label{dilaton-BH-2-b2r}
\eea
In terms of the Schwarzschild-like coordinate $t$, the right side of the TFD evolves forward and the left side evolves backward, as indicated by arrows in Fig.~\ref{fig:EE-2sided}. This setup generalizes the eternal black hole version of the information paradox discussed in \cite{Almheiri193,Almheiri194,Penington192} and in section~\ref{sec:1xBH} of the present paper to the case of two eternal black holes instead of one. Correspondingly, at late times we can expect islands to appear in both black hole regions. Similar models were considered in \cite{Penington192, Geng20, Geng21}.  See \cite{Anderson:2021vof, Balasubramanian:2021wgd} for a discussion of entanglement between disjoint gravitating universes.

To compute the generalized entropy functional, we again assume that the matter CFT is holographic and dual to some asymptotically AdS$_3$ geometry, and we follow the approach explained in Sec.~\ref{sec:1xBH-setup}. The CFT entanglement entropy can be computed by (\ref{EE-CFT}) using the conformal transformation to the plane from Sec.~\ref{sec:BHphase}. This transformation maps a pair of black hole exterior regions connected by a reservoir on the same side of the thermofield double (in Euclidean signature) to the complex $w$-plane with the conformally flat metric $\Omega^{-2}(w, \bar{w}) dw d\bar{w}$, with the Weyl factor $\Omega$ in the three regions given by (\ref{Omega1})-(\ref{Omegabath}): 
\bea
\text{Black hole 1}: &&\ \Omega_1 = \frac{1-|w|^2}{2}\,; \\
\text{Black hole 2}: &&\ \Omega_2 = \frac{|w|^2 - \left(\frac{\beta_2}{\beta_1}\right)^{\frac{2}{\alpha}}}{2 \left(\frac{\beta_2}{\beta_1}\right)^{\frac{1}{\alpha}}}\,; \\
\text{Reservoir}: &&\ \Omega_{\text{bath}} = \frac{2\pi}{\beta_1} |w|^{1-\alpha}\,.
\eea
Finally, to compute the area terms in the island formula (\ref{IslandFormula}) we need also the dilaton profile, which is given by equations (\ref{dilaton-BH-1-b1})-(\ref{dilaton-BH-2-b2}).

%%%%%%%%%%%%%%%
\subsection{Entanglement entropy of segments including the second black hole}
\label{sec:2xBH+BH}

\begin{figure}[t]
	\centering
	\includegraphics[scale=0.8]{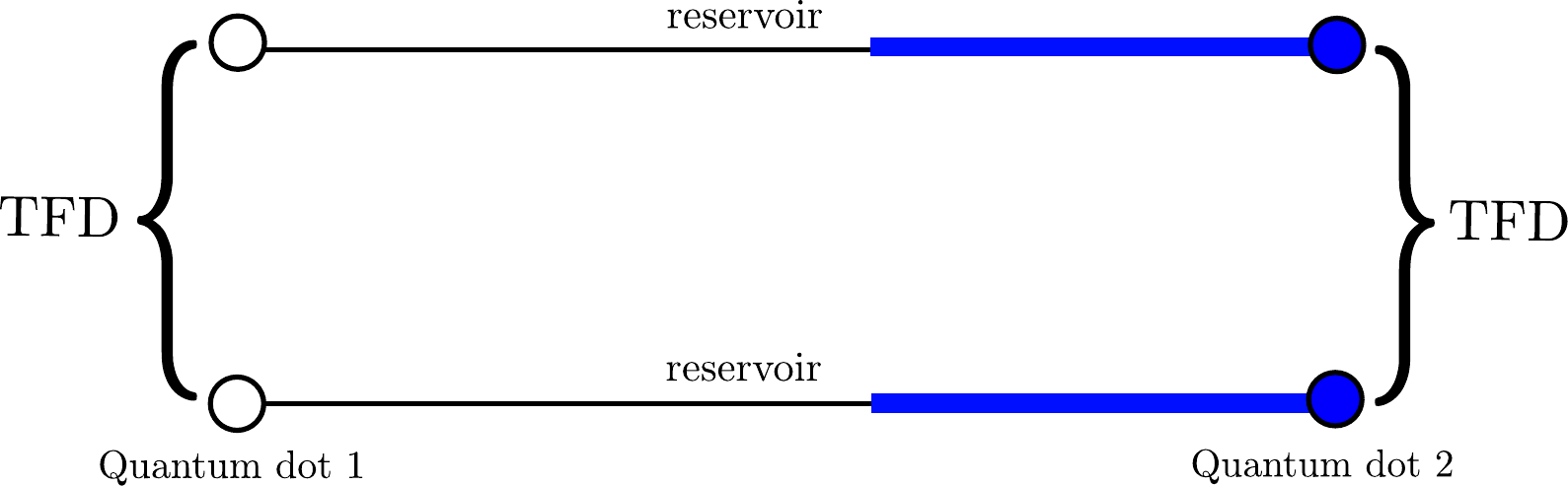} 
	\caption{We collect the radiation in the bold blue segments of radiation that include the boundary duals (quantum dots) of the second eternal black hole.}
	\label{fig:EE-2sided+BH}
\end{figure}
We begin by treating the second black hole as a detector which collects the radiation from the first black hole. This detector also radiates back into the reservoir, at a rate which creates an equilibrium with the reservoir heat engine. The Hilbert space  available to the black hole detector is finite, but nevertheless it is large enough to be comparable to the Hilbert space of the first black hole. This means that it should be able to gain access to the island.

\begin{figure}[t]
	\centering
	\includegraphics[scale=0.35]{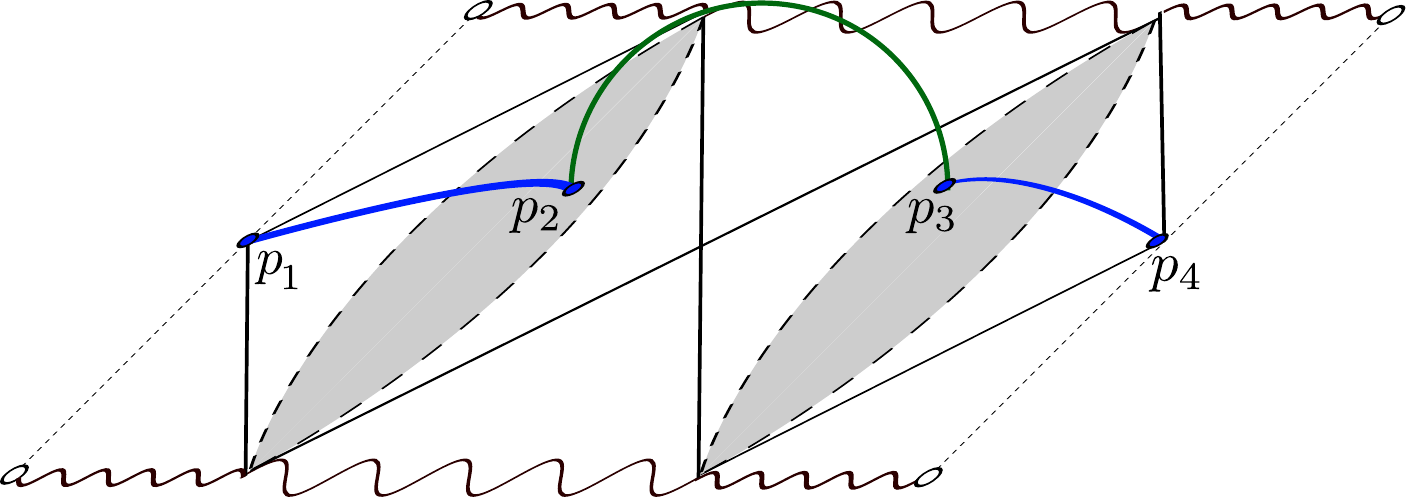}  
	\includegraphics[scale=0.35]{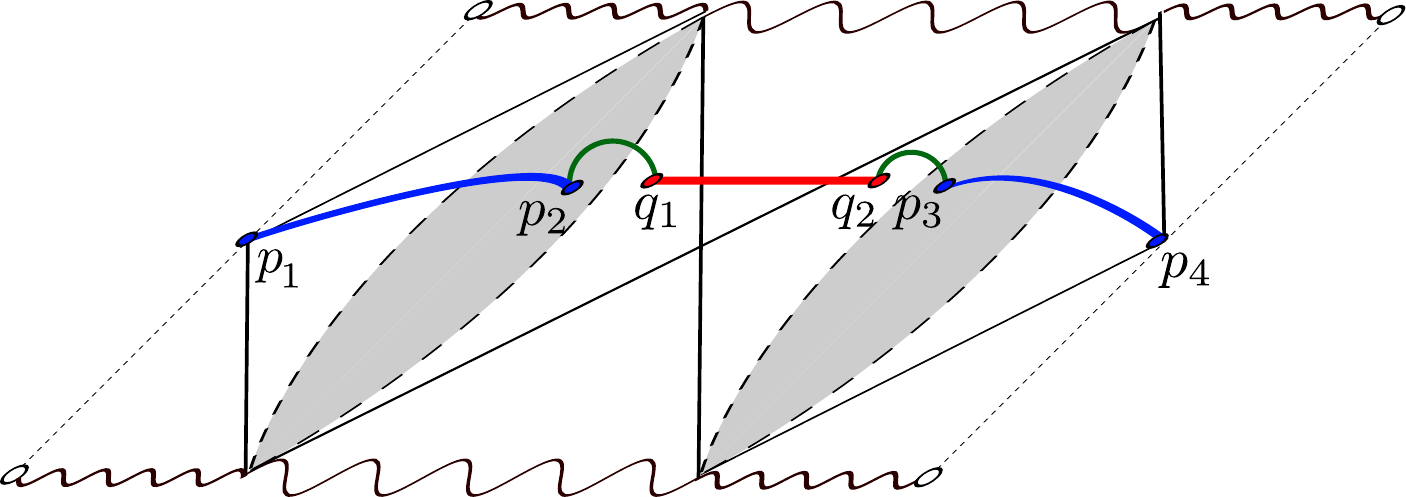} 
	\includegraphics[scale=0.35]{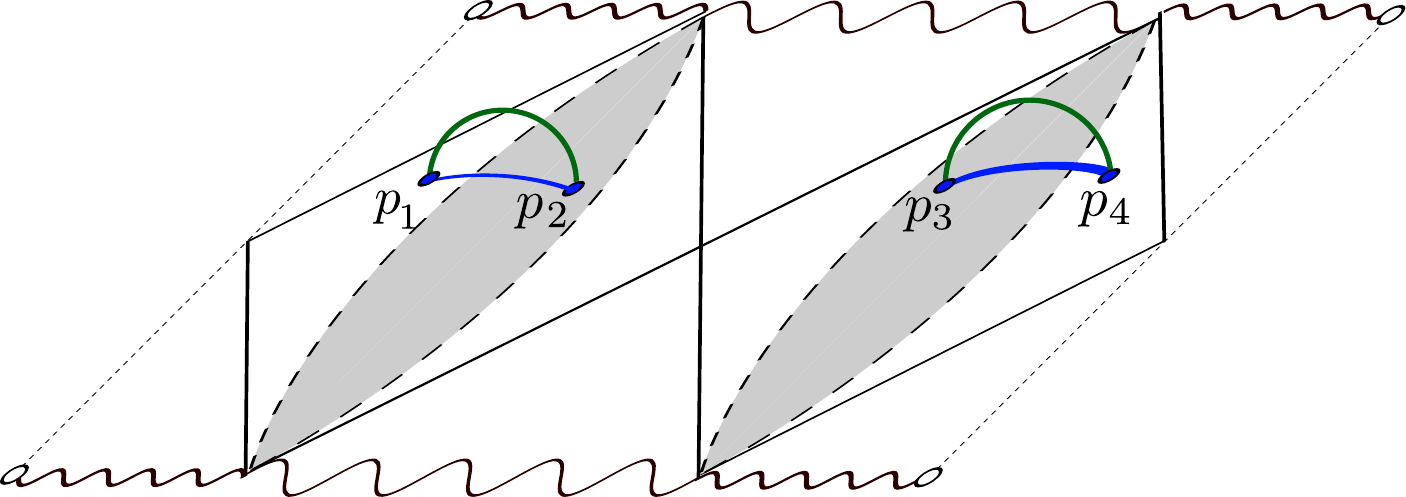} \\
	(a) \hspace{4cm} (b) \hspace{4cm} (c) \\ 
	\caption{Configurations of the RT geodesics extending into the 3d bulk which determine the competing generalized entropy channels with the second black hole included in the region being probed. The drawings are identified across the dashed line which goes through the bifurcation surface of the second black hole. Multiple-island configurations, which turn out to be always subleading, are not shown. For the connected configurations (a) and (b) the blue segments join across the bifurcation surface and the QES is empty surface.}
	\label{fig:EE-2sided+BH-geod}
\end{figure}

In the microscopic description, we want to compute the entanglement entropy of a quantum dot and its thermofield double partner, plus some of their adjoining baths (Fig.~\ref{fig:EE-2sided+BH}). The coordinates of the reservoir endpoints in terms of the $\xi$, $t$ variable are chosen as 
\be
 p_2 = (a, -t+i\pi); \quad p_3 = (a, t)\,.
\ee 
In the two-dimensional effective gravitational description we need to search for QESs which extend the regions into the bulk, plus possible QESs in the other gravitational region which bound islands. Three configurations\footnote{We omit the usual UV divergences in the generalized entropy functionals and other formulae for the entanglement entropy as we did in Sec.~\ref{sec:1xBH}.}  can dominate  the Page curve; they are shown in Fig.~\ref{fig:EE-2sided+BH-geod} (other configurations are subleading). As we assume the CFT is holographic, it is straightforward to identify the channels in the CFT entanglement entropy which define the competing generalized entropy functionals and quantum extremal island configurations.
%Let us discuss them in detail. 

\paragraph{Configuration (a): linear growth.} This is a fully connected no-island configuration, with a trivial (empty) QES in the second black hole region. The endpoints $p_2 = (a, -t+i \pi)$ and $p_3 = (a, t)$ are connected by a geodesic in the 3D bulk. With this in mind, the entanglement entropy of the configuration (a) is given just by one geodesic shown in Fig.~\ref{fig:EE-2sided+BH-geod}(a), and reads 
\bea
\mS_a = S_{\text{conn.}}^{\text{no island}} (p_2, p_3) = \frac{c}{3}\log \left( \frac{\beta_1 e^{ \alpha a }}{\pi} \cosh t\right)\,.
\eea

\paragraph{Configuration (b): the island.} This is a fully connected configuration that includes an island in the first black hole region and  an empty QES in the second black hole region. The location of the island $[q_{1}, q_{2}]$ is again defined by the QESs. Since the points $p_2$ and $p_3$ are located symmetrically with respect to the bifurcation surface of the first black hole, that means that the QES points will have the same coordinates $q_{1,2} = (x, t_x)$. The generalized entropy is
\be
\mS_b = \underset{q_1}{\Ext}\ S_{\text{gen}}^{\text{island}} (q_1, p_2) + \underset{q_2}{\Ext}\ S_{\text{gen}}^{\text{island}} (q_2, p_3) \,,
\ee
where the generalized entropy functional reads
\bea
S_{\text{gen}}^{\text{island}} (q_2, p_3) &=&  \phi_0 + \frac{2\pi \phi_r}{\beta_1} \coth \left(-x\right) 
\nn\\ &+&\frac{c}{6} \log \left(\frac{\beta_1 e^{\alpha a}\left(\cosh (x-a) - \cosh \left(t - t_x\right)\right)}{\pi \sinh\left(- x\right)}\right)\,. \label{Sgen1} 
\eea
Extremizing $t_x$ gives $t_x = t$, and so the island contribution is time-independent. 

\paragraph{Configuration (c): disconnected QES configuration.} This configuration corresponds to two times the entropy of the segment. The corresponding generalized entropy functional reads
\bea
S_{\text{gen}}(p_3, p_4) &=& \phi_0 + \frac{2\pi \phi_r}{\beta_2} \coth \left(y - \alpha^{-1}\log \frac{\beta_2}{\beta_1}\right) \nn\\ &+& \frac{c}{6} \log \left(\frac{2\beta_1 e^{\alpha a + y} \frac{\beta_2}{\beta_1}^{1/\alpha} \left[\cosh \left(y - a\right) - \cosh ( t - t_y)\right]}{\pi \left(e^{2y} - \left(\frac{\beta_2}{\beta_1}\right)^{2/\alpha}\right)}\right)\,, \label{Sgen2}
\eea
where $(y, t_y)$ are the coordinates of the QES $p_4$. The total entropy in this configuration reads 
\be
\mS_c = 2 \times \underset{p_4}{\Ext}\ S_{\text{gen}} (p_3, p_4) \,.
\ee
It is worth noting that for this configuration the QES points $p_{1,4}$ end up at finite distance between the AdS$_2$ boundary and the horizon in the corresponding external regions of the second black hole. This quantity is also time-independent.

\subsubsection{Page curves}
\begin{figure}[t]
	\centering
	\includegraphics[scale=0.4]{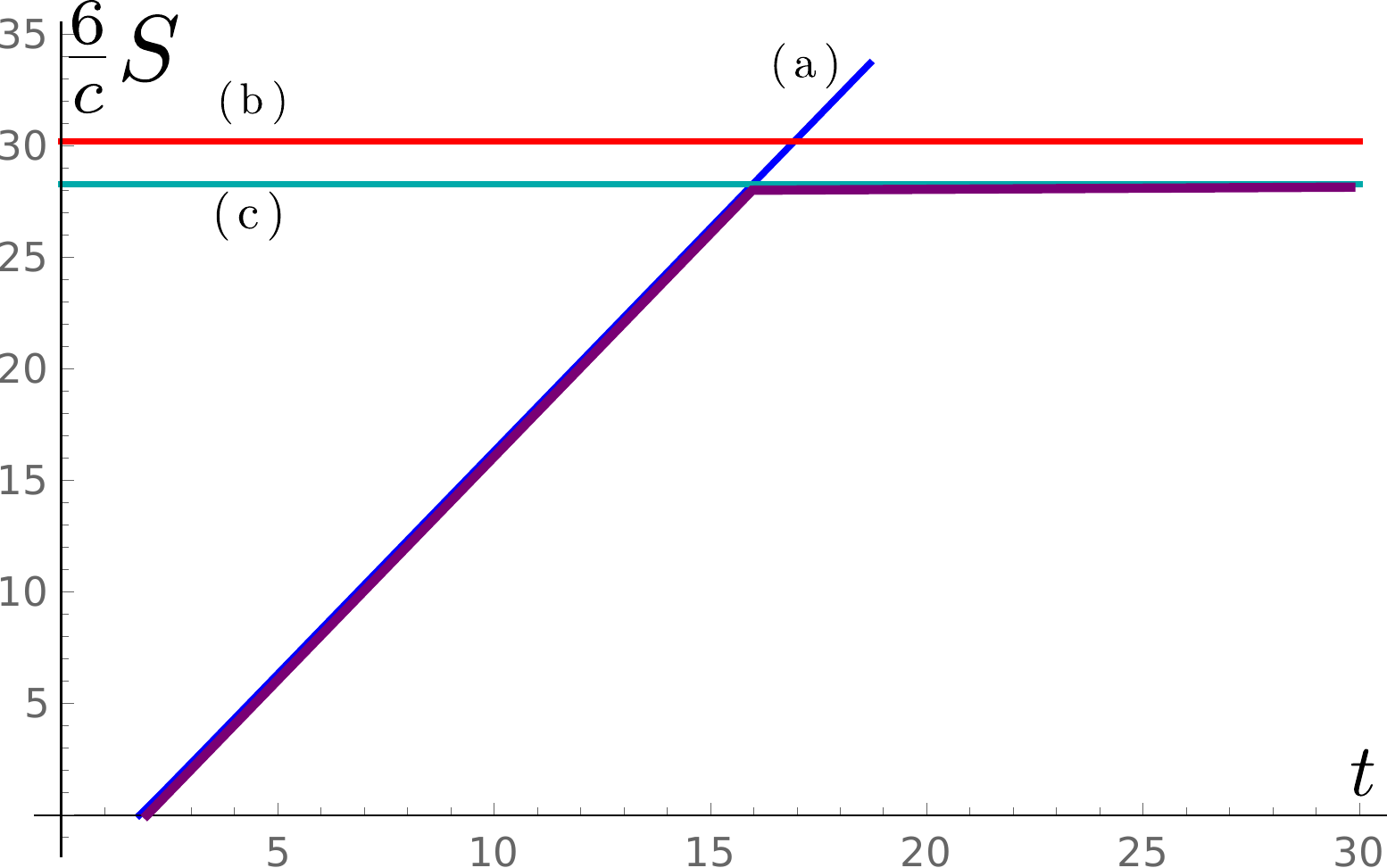} 
	\includegraphics[scale=0.4]{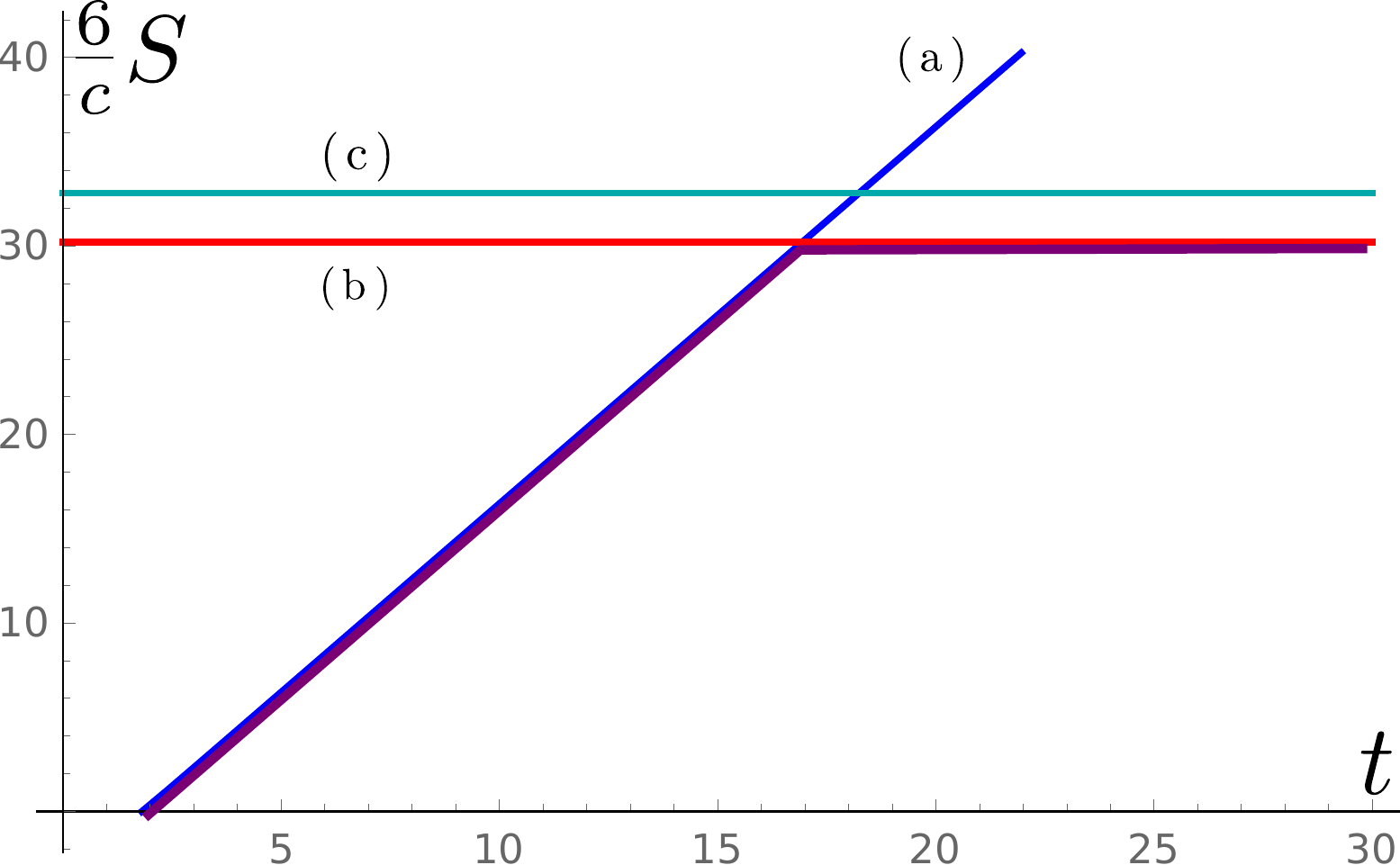} \\
	A \hspace{5cm} B 
	\caption{Page curves for the radiation segments plus the second black hole. We fix the physical reservoir size $L_R = 0.5$, physical location of $p_{2,3}$ endpoints $r = r_1 + 0.01 L_R$, $\beta_1 = 1$, $\phi_0 = 10\frac{c}{6}$ and $\phi_r = \frac{c}{6}$. (A) $\beta_2 = 2$. The no-island phase dominates at late times.  (B) $\beta_2 = 1.3$. The island phase dominates at late times. }
	\label{fig:Page-1+1BH}
\end{figure}
The possible behaviors in the case of the black hole detector involve a competition between the two phases of constant entanglement entropy at late times -- namely, between configurations (b) and (c). Let us fix the physical size of the reservoir $L_R$. Then we can have a transition between (b) and (c) at late times if we vary the temperature ratio $\beta_2 / \beta_1$ or the physical position of the points $p_{2,3}$ at the Rindler coordinate $r = \frac{\beta_1}{2\pi \alpha} e^{\alpha a}$. We show this transition in Fig.~\ref{fig:Page-1+1BH} for the case when we vary the temperature of the ``detector'' $\beta_2$ while keeping $r$ fixed. We see that adjusting this temperature can reveal the island of the first black hole. A similar results were obtained in \cite{Chen:2020jvn,Geng21}. 

This transition has a simple interpretation, if we think of the island configuration (b) as the entanglement entropy of the first black hole plus the segments $[0, a]$. It competes against configuration (c), which is the entropy of the second black hole plus segments $[a, L]$. Then the island of the first black hole reveals itself if the entropy of the first black hole becomes smaller than the entanglement entropy of the probed system (which includes the second black hole and the radiation segment $[a, L]$). This means that this system, which we can think of as a detector, has enough room to effectively accommodate all states of the radiation. If the detector's entropy is not large enough, however, it effectively thermalizes before gaining access to the island, which is expressed by the phase (c). In other words, the no-island phase 9(c)  and the island phase 9(b) can be interpreted, respectively, as the island phase and the no-island phase of the computation of entanglement entropy of the complementary subsystem. Note that the key property in the evolution of the entanglement entropy of radiation regions which include one of the black holes is that the Hilbert space of the radiation is large. Because of this, there are no finite size effects that would introduce an intermediate phase into the Page curve for any  choice of parameters, as we saw in the case of a single black hole in Sec.~\ref{sec:1xBH}. Below, we  exclude both black holes from the radiation region, so that the radiation segment belongs to the interior of the reservoir only. We will then see that the Page curve structure becomes richer. 

%%%%%%%%%%%%%
\subsection{Entanglement entropy of segments in reservoir}
\label{sec:2xBH-bath}

\begin{figure}[t]
	\centering
	\includegraphics[scale=0.8]{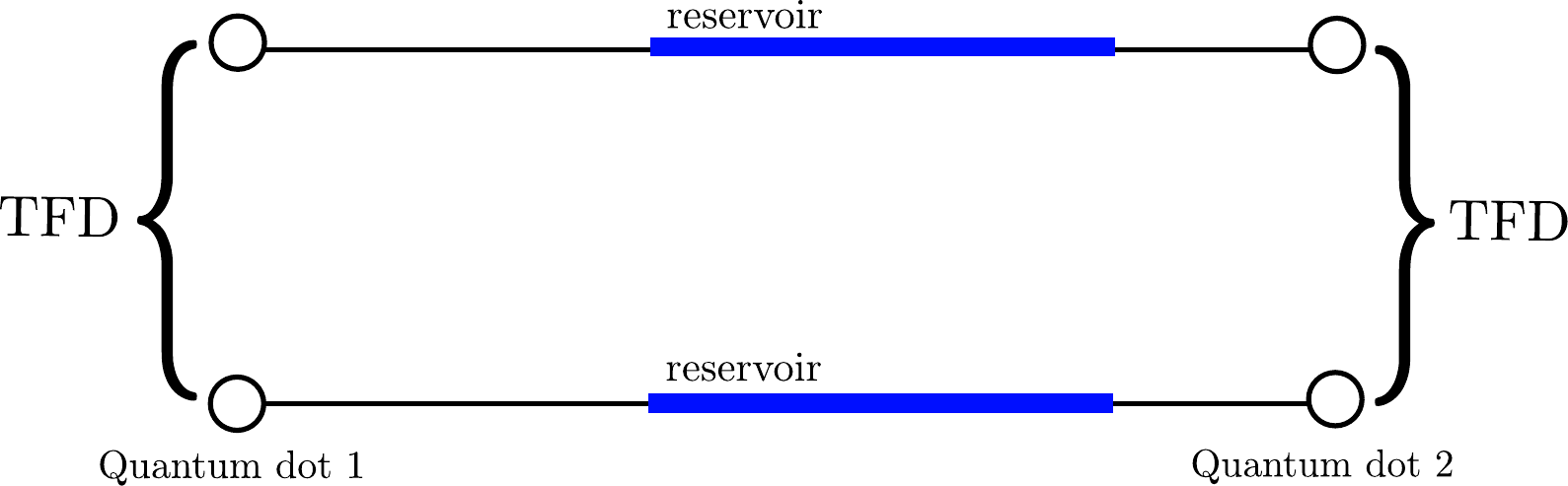} 
	\caption{We consider the radiation in the bold blue segments of the reservoir.}
	\label{fig:EE-2sided-bath}
\end{figure}

\begin{figure}[t]
	\centering
	\includegraphics[scale=0.35]{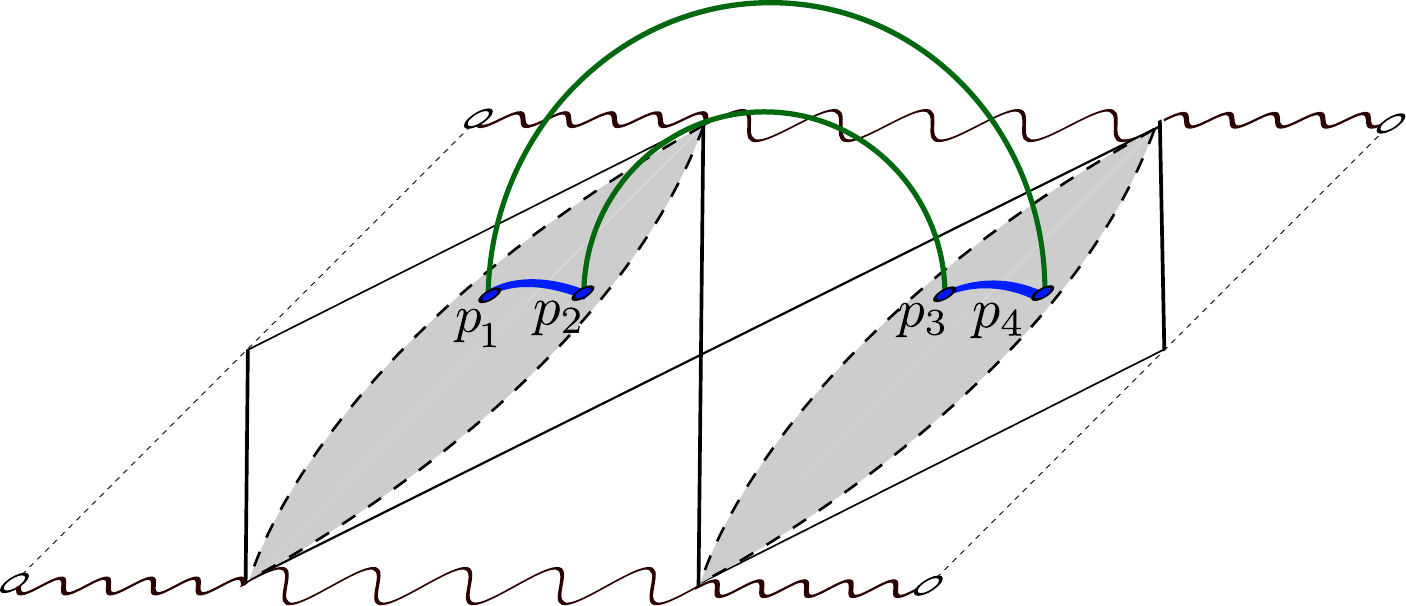}  
	\includegraphics[scale=0.35]{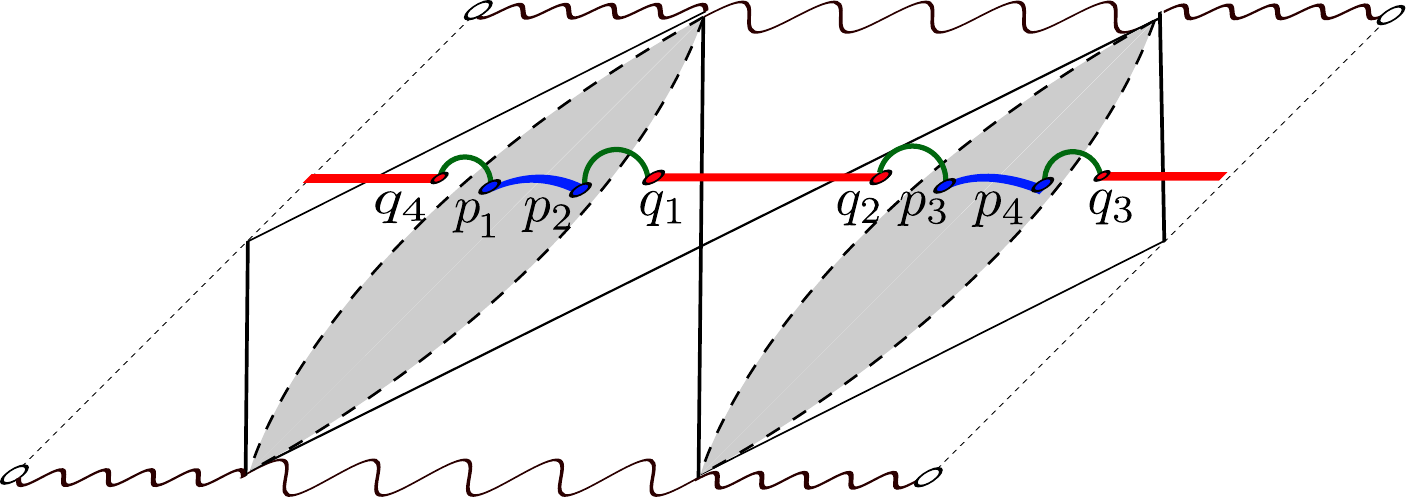} 
	\includegraphics[scale=0.35]{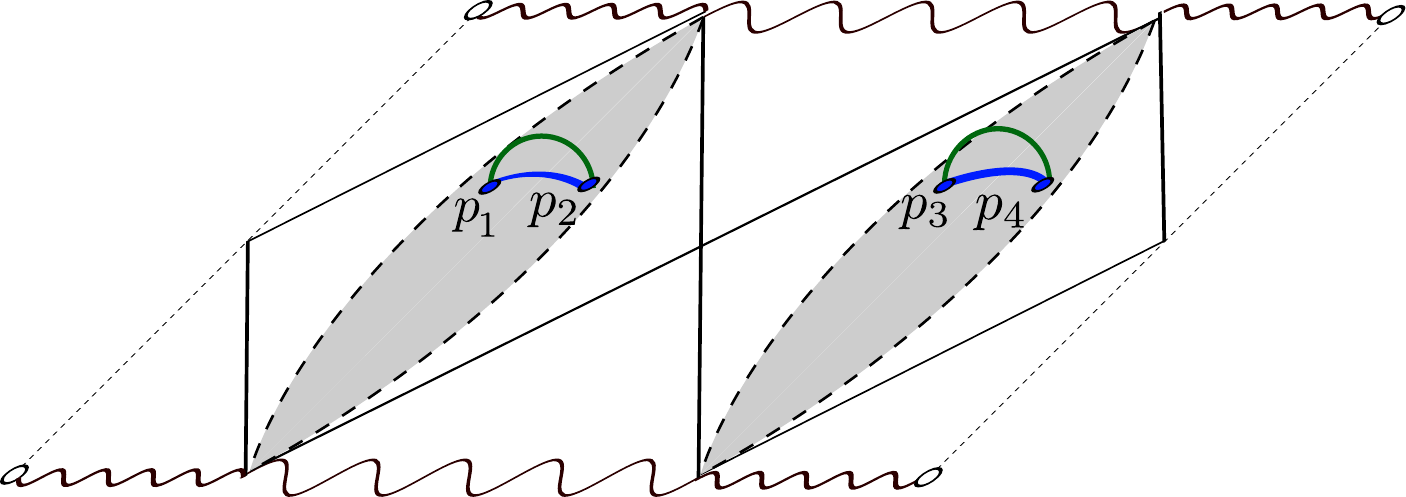} \\
	(a) \hspace{4cm} (b) \hspace{4cm} (c) \\ 
	\includegraphics[scale=0.35]{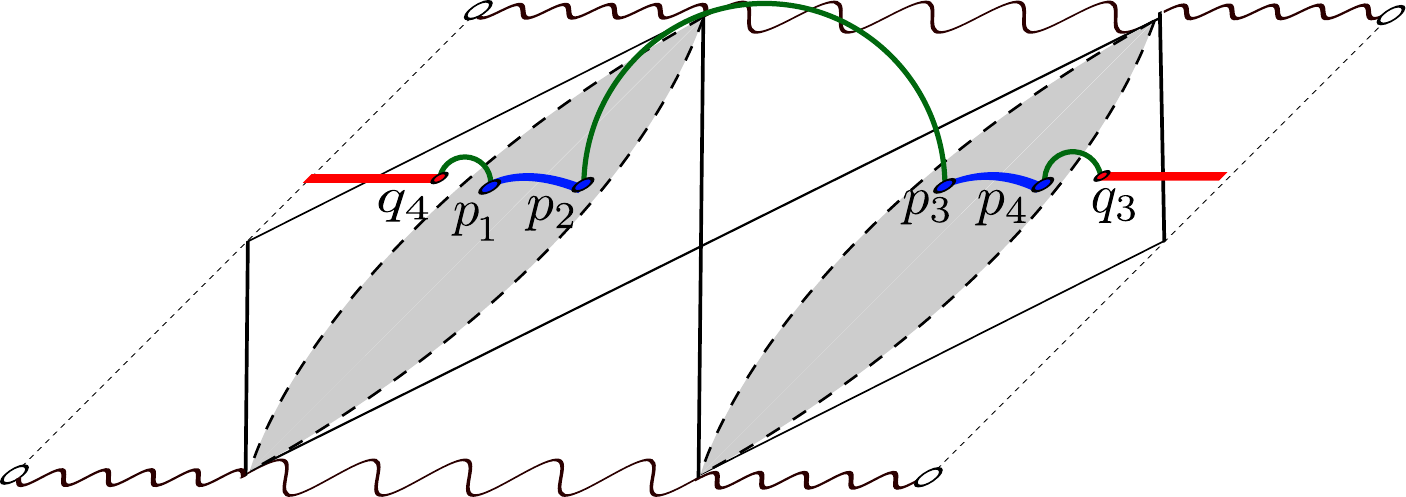} 
	\includegraphics[scale=0.35]{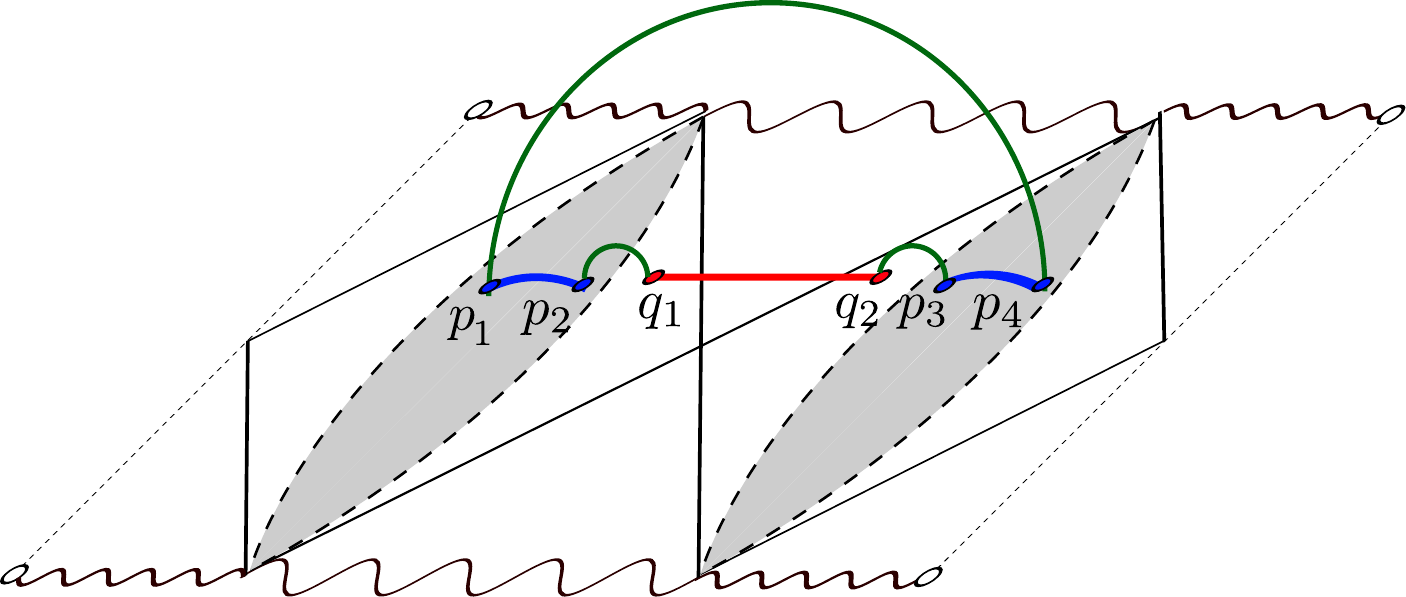}\\ 
	(d) \hspace{4cm} (e)
	\caption{Configurations of the RT geodesics extending into the 3d bulk which determine the competing generalized entropy channels. The drawings are identified across the dashed line which goes through the bifurcation surface of the second black hole. Multiple-island configurations that are always subleading are not shown.}
	\label{fig:EE-2sided-bath-geod}
\end{figure}

We compute the entanglement entropy of the union of two identical segments $A = [p_1, p_2] \cup [p_3, p_4]$ positioned inside corresponding copies of the reservoir, as shown in Fig.~\ref{fig:EE-2sided-bath}. The coordinates of these endpoints are chosen as: 
\be
p_1 = (b, -t+i\pi); \quad p_2 = (a, -t+i\pi); \quad p_3 = (a, t); \quad p_4 = (b, t)\,.
\ee
We now have 5 possible generalized entropy configurations which can dominate, shown in Fig.~\ref{fig:EE-2sided-bath-geod}.

\paragraph{Configuration (a): linear growth.} This is a fully connected no-island configuration, where two RT geodesics connect the endpoints $p_1 \leftrightarrow p_4$ and $p_2 \leftrightarrow p_3$, respectively. The entanglement entropy of this configuration is given by the formula 
\bea
\mS_a = S_{\text{conn.}}^{\text{no island}} (p_1, p_2; p_3, p_4) = 2\frac{c}{3}\log \left( \frac{\beta_1 e^{ \frac{\alpha (a+b)}{2}} }{\pi} \cosh t\right)\,.
\eea
This expression grows in time approximately linearly, and in the general case the entropy depends on the segment location when $\alpha \neq 0$. 

\paragraph{Configuration (b): two islands.} This is a fully connected configuration that includes an island in every black hole region. The location of the islands $[q_{1,3}, q_{2,4}]$ is defined by the QESs, obtained by extremizing the generalized entropy functionals. These generalized entropy functionals are exactly the same ones as (\ref{Sgen1}) for the island in the first black hole and (\ref{Sgen2}) for the island in the second black hole. We write down the corresponding formulae for the QESs on the right side of the thermofield double, the points $q_2 = (x, t_x)$ and $q_3 = (y, t_y)$. The QESs $q_1$ and $q_4$ on the left are determined analogously. 
\bea
S_{\text{gen}}^{\text{island}} (q_2, p_3) &=&  \phi_0 + \frac{2\pi \phi_r}{\beta_1} \coth \left(- x\right) 
\nn\\ &+&\frac{c}{6} \log \left(\frac{\beta_1 e^{\alpha a}\left(\cosh(x-a ) - \cosh \left(t -  t_x\right)\right)}{\pi \sinh\left(-x\right)}\right)\,; \label{Sgen11} \\
S_{\text{gen}}^{\text{island}} (q_3, p_4) &=& \phi_0 + \frac{2\pi \phi_r}{\beta_2} \coth \left(y - \alpha^{-1}\log \frac{\beta_2}{\beta_1}\right) \nn\\ &+& \frac{c}{6} \log \left(\frac{2\beta_1 e^{\alpha b + y} \frac{\beta_2}{\beta_1}^{1/\alpha} \left[\cosh \left(y- b\right) - \cosh \left( t -  t_y\right)\right]}{\pi \left(e^{y} - \left(\frac{\beta_2}{\beta_1}\right)^{2/\alpha}\right)}\right)\,. \label{Sgen22}
\eea
The total entropy in this configuration is given by the sum of the two island contributions, or four quantum extremal surfaces: 
\be
\mS_b = \underset{q_2}{\Ext}\ S_{\text{gen}}^{\text{island}} (q_2, p_3) + \underset{q_3}{\Ext}\ S_{\text{gen}}^{\text{island}} (q_3, p_4) + \underset{q_4}{\Ext}\ S_{\text{gen}}^{\text{island}} (q_4, p_1) + \underset{q_1}{\Ext}\  S_{\text{gen}}^{\text{island}} (q_1, p_2)\,.
\ee
The two-island contribution is time-independent, and $t_x = t_y = t$.

\paragraph{Configuration (c): generalized thermalization.} This is a disconnected configuration, which is the sum of the CFT entanglement entropy for the two thermofield double copies of the reservoir segment. The entanglement entropy of a copy is given by the formula
\be
S_{\text{disc.}}(p_1, p_2) = \frac{c}{3} \log\left( \frac{\beta_1 e^{\alpha(a+b)} }{\pi} \sinh \frac{|a - b|}{2}\right)\,. \label{Sdisc}
\ee
This configuration can be interpreted as describing the extension of the notion of thermalization to the case of the varying temperature. The entropy of the segment on the left of TFD $[p_3, p_4]$ is given by an analogous formula, and the total entropy in this configuration is given by 
\be
\mS_c = S_{\text{disc.}}(p_1, p_2) + S_{\text{disc.}}(p_3, p_4)\,.
\ee

\paragraph{Configurations (d) and (e): single island.}

In these configurations we have an island in one of the black holes while the other black hole still produces linear growth: 
\bea
\mS_d &=& \underset{q_4}{\Ext}\ S_{\text{gen}}^{\text{island}} (q_4, p_1) + \underset{q_3}{\Ext}\ S_{\text{gen}}^{\text{island}} (q_3, p_4) + \frac{c}{3}\log \left( \frac{\beta_1 e^{a \alpha } }{\pi} \cosh t\right) \,;\\
\mS_e &=& \underset{q_1}{\Ext}\ S_{\text{gen}}^{\text{island}} (q_1, p_2) + \underset{q_2}{\Ext}\ S_{\text{gen}}^{\text{island}} (q_2, p_3) + \frac{c}{3}\log \left( \frac{\beta_1 e^{\alpha b} }{\pi} \cosh  t \right)\,.
\eea
These expressions grow linearly with half the slope of configuration (a). 

Let us note that there are also configurations that have two islands in black hole regions, similarly to the case (d) in Sec.~\ref{sec:1xBH} (see Fig.~\ref{fig:1BH-geod}). Just as in that case of a single eternal black hole, such configurations are always subleading, so we do not consider them. 

\subsubsection{Page curves}

\begin{figure}[t]
	\centering
	\includegraphics[scale=0.4]{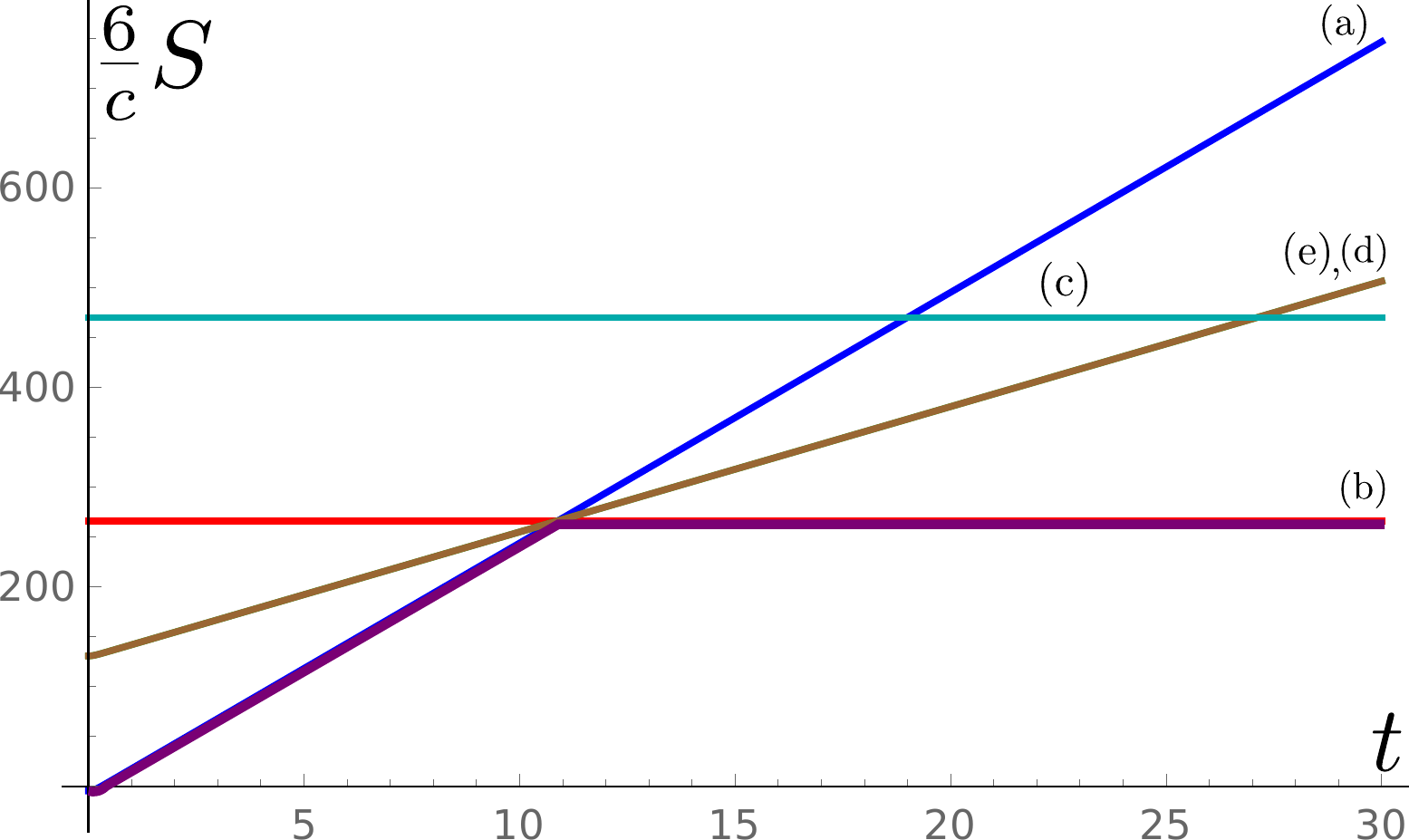} 
	\includegraphics[scale=0.4]{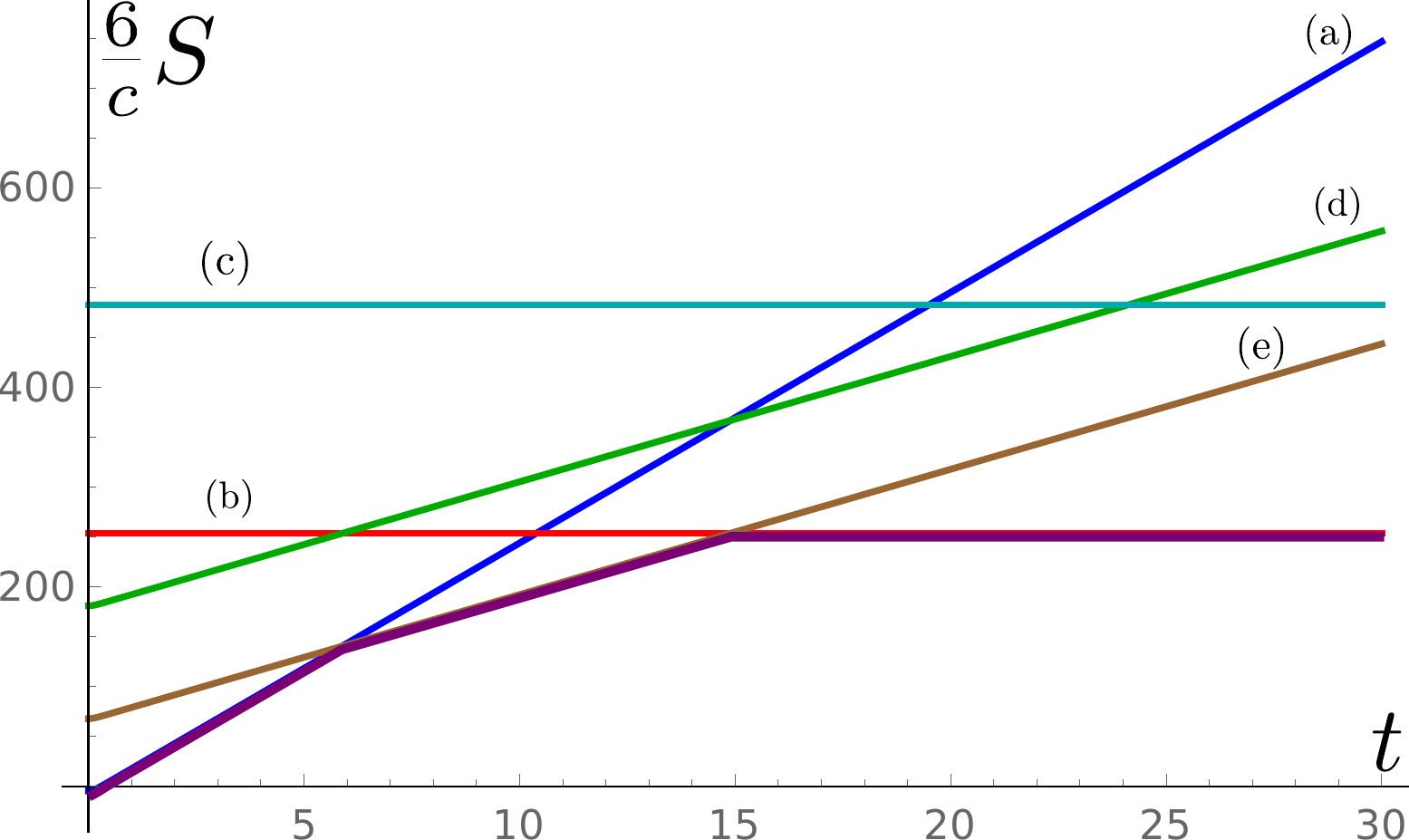} \\
	A \hspace{5cm} B \\
	\includegraphics[scale=0.4]{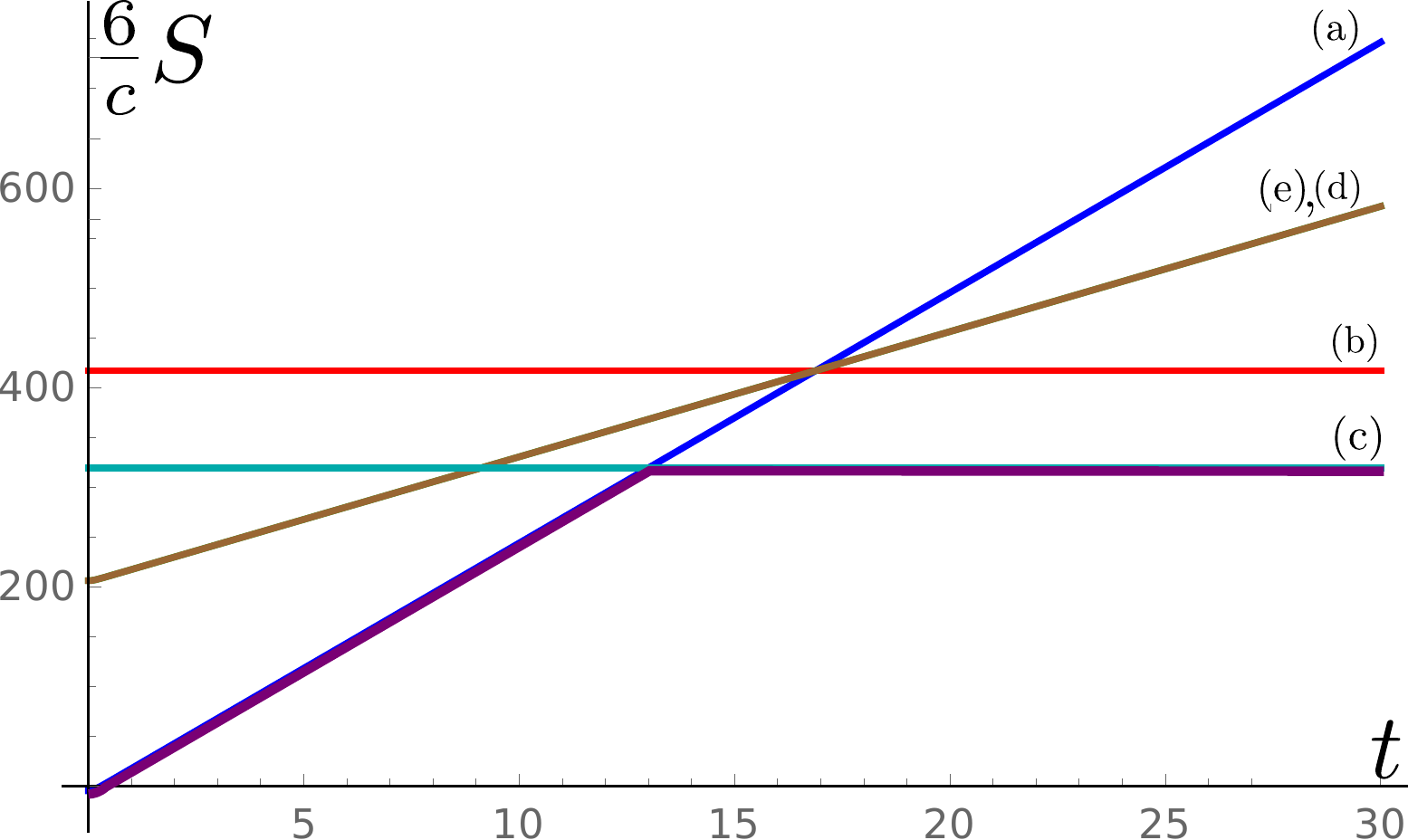} 
	\includegraphics[scale=0.4]{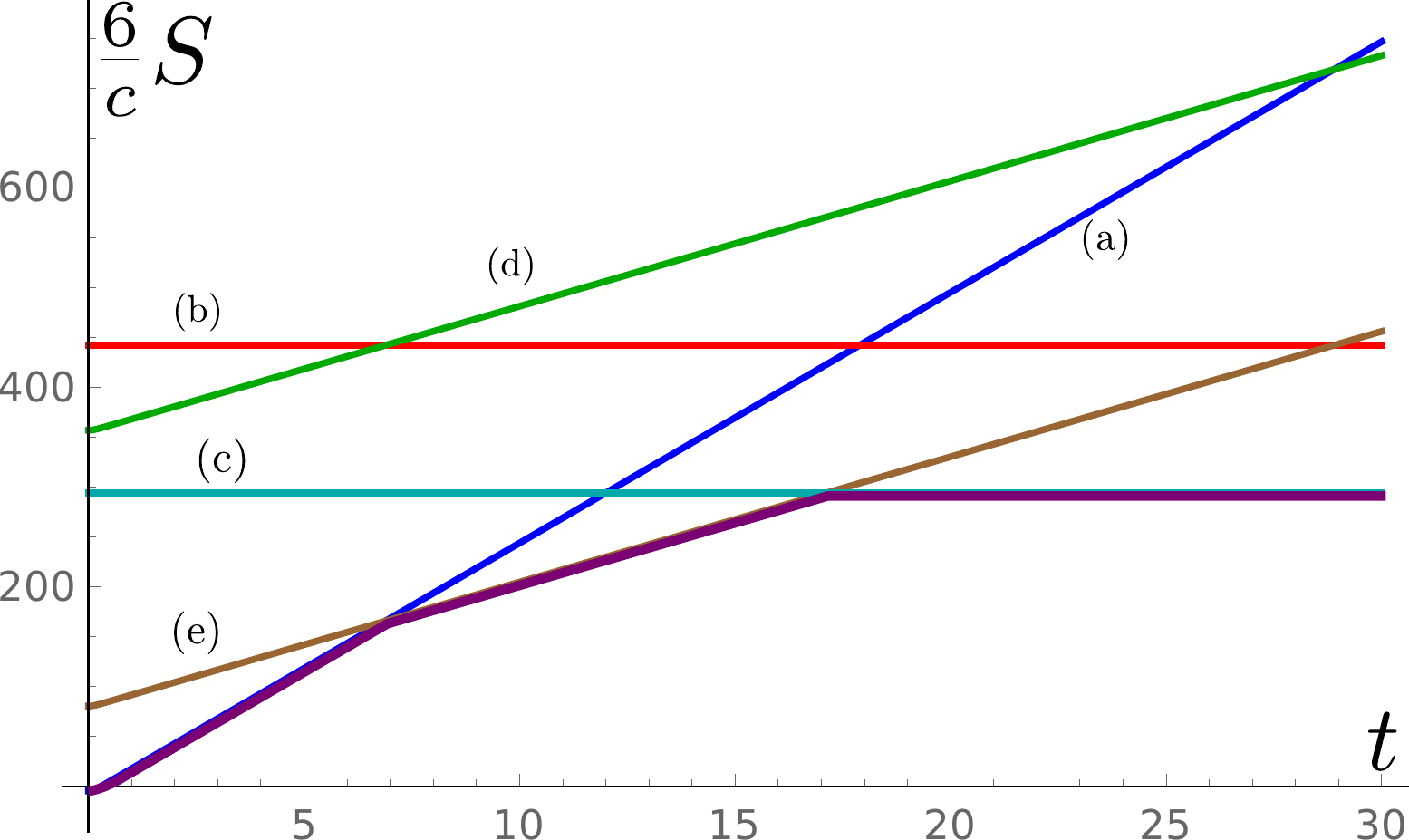} \\
	C \hspace{5cm} D
	\caption{Page curves for the radiation segments in the reservoir in the equal temperature case with $\beta_1 = \beta_2 = 1$. (A) $a= 0.12 L$, $b = 0.88 L$. (B) $a= 0.02 L$, $b = 0.78 L$. (C) $a= 0.24 L$, $b = 0.76 L$. (D) $a= 0.04 L$, $b = 0.52 L$. We have set $\frac{c}{6}\phi_0 = 20$ and $\frac{\pi \phi_r}{3} = 10$.}
	\label{fig:Page-bath-equal}
\end{figure}
We have four qualitatively different possibilities for the Page curve, which are determined by the competition between the two-island phase (b) and the disconnected phase (c) at very late times, and by the possibility of one of the two single-island phases (d) or (e) dominating for a finite time betwe
en the early and very late times. 

First, let us consider the equal-temperature case, which is achieved in the limit $\alpha\to 0$, $\frac{\beta_2}{\beta_1} \to 1$ with $L = 2\pi L_R/\b$ fixed. The four qualitatively different Page curves are shown in Fig.~\ref{fig:Page-bath-equal}. More specifically, we plot the contributions of the five phases (a) -- (e) described above. 
\begin{itemize}
    \item In Fig.~\ref{fig:Page-bath-equal}A the linear growth generated by the connected no-island phase (a) transitions into the two-island phase (b), which keeps the entanglement entropy constant. The segment is large enough so that the island phase saturates the entanglement entropy before the disconnected phase (c) has a chance to become relevant. 
    \item In Fig.~\ref{fig:Page-bath-equal}B the connected phase (a) also dominates at early time, but because of the de-centered position of the segment in the reservoir, at some time the transition to the mixed phase (e) happens. This phase has an island in the first black hole but continues to grow linearly with halved slope due to the active remaining ER bridge across the second black hole. At later times, another transition happens, where the entanglement entropy is completely saturated by the two-island phase. Thus, the single-island phase persists only for finite time. 
    \item In Fig.~\ref{fig:Page-bath-equal}C, the segment is small enough so that the entanglement entropy is saturated by the disconnected phase (c) before any islands get involved. 
    \item In Fig.~\ref{fig:Page-bath-equal}D, the early-time growth (a) transitions into the mixed phase (e) with slower growth and an island in one of the black holes. However, later the entanglement entropy is saturated by the disconnected phase (c) again, so the island is no longer accessible. Thus, this behavior only allows for temporary access to an island. 
\end{itemize}
When the temperatures are different, $\beta_1 \neq \beta_2$, the qualitative behavior is the same. The main difference is that now  temporary single-island phases are possible even for  segments centered in the reservoir, because there is an additional non-uniform contribution in the entanglement entropy arising from the heat engine, which is represented, e.g., by the second term in (\ref{Sdisc}). Because of this, a Page curve for any fixed pair of segments can include a temporary single island phase if we adjust the temperature difference and/or the reservoir size and positions of endpoints of $A$. 

By utilizing the unitarity bound $S(A) \leq$ min$\{\log \dim \mathcal{H}_A, \log \dim \mathcal{H}_{\bar{A}}\}$ discussed in Sec.~\ref{intro} and applied at the end of Sec.~\ref{sec:1xBH}, the qualitative nature of these transitions can be predicted. In our setting, both $\dim  \mathcal{H}_A$ and $ \dim \mathcal{H}_{\bar{A}}$ are finite, and so these Hilbert spaces enforce distinct unitarity bounds. When the former is threatened, there is a thermalization transition (c), whereas when the latter is threatened, we are led to the two-island phase (b). There is also the possibility of transition to one-island phases (d) or (e), which have growing entropy and therefore are not explained by the simple unitarity bound. As in Sec.~\ref{sec:1xBH}, this case can be explained by a unitarity bound applied to the two gravitational regions (plus the adjoining baths up to the radiation region $A$) separately. When the entropy of this region approaches the entropy of the two black holes in the given gravitational region $2 S_{BH}$, a transition must occur. 

%%%%%%%%%%%%%%%%%%%%%%%%%%%%%%%%%%%%%%%%%%%%%%%%%%%%%%%%%%%%%%%%%%%%%%%%%%%%%
\section{Discussion}

We have investigated the effects of  competition between the thermalization mechanism and the island information recovery mechanism in the time evolution of entanglement entropy of Hawking radiation. We considered finite radiation regions in the classic example of a thermofield double black hole coupled to two semi-infinite baths. We also introduced a new model, where we have two pairs of thermofield double black holes at different temperatures  radiating into a finite shared bath (and its thermofield double). To maintain equilibrium this required a Rindler-like bath which operated as a heat engine equilibrating the two sides. A summary of our results can be found in Sec.~\ref{intro}.

The two-temperature system provides a tunable parameter which models thermal loss from engineered quantum dots in the lab, such as the SYK system. In relation to this, it should also be possible to write down an explicit coupling between  two SYK systems at different temperatures  leading to asymmetric wormholes, generalizing the Maldacena-Qi solution \cite{Maldacena18}.\footnote{There are avatars of the asymmetric wormhole in contexts where it is not the leading saddle of a path integral; for example in the matrix model description of JT gravity \cite{SSS,SW} it is an off-shell configuration captured by $\langle Z(\b_1) Z(\b_2)\rangle$, while in systems with multiple uncoupled SYK dots there may exist subleading saddles which link the various systems (such subleading saddles have been exhibited for the case of equal temperatures in \cite{AKTV}).}

It would be interesting to study the generalization of our two-temperature model to higher dimensions. In two dimensions we saw that we can have transitions from the asymmetric wormhole (the ``confined" phase) to  two thermofield double black holes (the ``deconfined" phase). In higher dimensions, by picking the bath and boundary CFTs appropriately, one can more cleanly probe the confining structure of the model as the temperatures are varied through an order parameter like center symmetry \cite{Witten:1998zw}, which does not have much meaning in $(0+1)$ dimensions. In the two-temperature model considered in this paper, the fate of one of the thermofield double quantum dots was tied to the other one, since the confining phase was a gravitational solution which linked the two systems. In higher dimensions, however, the confining phase of a single thermofield double boundary pair is two copies of thermal AdS, and it need not link to the other thermofield double. Thus we can have one of the thermofield double boundary pairs in the confined phase while the other pair is in the deconfined phase. It would be interesting to work out whether this actually occurs, and, if so, what the doubly holographic solution looks like; it would need to have nontrivial topology to accommodate the change from confined to deconfined phases.

The computations of entanglement entropy in the black hole phase would also take on a different structure in higher dimensions. The closest analogy to our results would be obtained by taking the bath CFT$_d$ to be placed on $\mathbb{R}^d$ and considering infinite strips, which would allow the ``halfway" surfaces seen in two dimensions, as in Fig.~\ref{fig:1BH-geod}(b). Such surfaces would not appear for compact regions. One way to see this is to note that in Fig.~\ref{fig:1BH-geod}(b) the two endpoints of an interval are treated differently, which would not work for the boundary of a higher-dimensional ball, which is connected.

%%%%%%%%%%%%%%%%%%%%%%%%%%%%%%%%%%%%%%%%%%%%%%%%%%%%%%%%%%%%%%%%%%%%%%%%%%%%
\section*{Acknowledgments}

The authors would like to thank Raghu Mahajan for valuable discussions. This work has been supported in part by FWO-Vlaanderen through project G006918N and by Vrije Universiteit Brussel through the Strategic Research Program ``High-Energy Physics.''
Research at the University of Pennsylvania was supported by the Simons Foundation It From Qubit collaboration (385592) and the DOE QuantISED grant DESC0020360. VB thanks the Santa Fe Institute for hospitality as this work was completed. The work of M.~K. was funded by Russian Federation represented by the Ministry of Science and Higher Education (grant number 075-15-2020-788).

%%%%%%%%%%%%%%%%%%%%%%%%%%%%%%%%%%%%%%%%%%%%%%%%%%%%%%%%%%%%%%%%%%%%%%%%%%
\appendix

%%%%%%%%%%%%%%%%%%%%%%%%%%%%%%%%%%%%%%%%%%%%%%%%%%%%%%%%%%%%%%%%%%%%%%%%%%
\bibliographystyle{JHEP}
\bibliography{2BH.bib}

\begin{comment}

\end{comment}

%\footnotesize
%\bibliographystyle{apsrev4-1long}
%\bibliography{babybib}

\end{document}